\documentclass[prl,twocolumn,notitlepage,longbibliography]{revtex4-1}
\usepackage{amsmath}
\usepackage{amssymb}

\usepackage[unicode=true,colorlinks=true,citecolor=blue,urlcolor=blue]{hyperref}

\usepackage{color}

\usepackage[normalem]{ulem}
\usepackage{bm}
\usepackage{epsfig}

\usepackage[normalem]{ulem}

\renewcommand {\Im}{\mathop\mathrm{Im}\nolimits}
\renewcommand {\Re}{\mathop\mathrm{Re}\nolimits}

\renewcommand {\i}{{\rm i}}
\renewcommand {\phi}{{\varphi}}

\begin{document}
\title{Stable interaction-induced Anderson-like localization embedded in standing waves}

\author{Na Zhang$^{1,2,3}$}

\author{Yongguan Ke$^{1,2}$}
\altaffiliation{Email: keyg@mail2.sysu.edu.cn}
\author{Ling Lin$^{2}$}
\author{Li Zhang$^{2}$}
\author{Chaohong Lee$^{1,2,3}$}
\altaffiliation{Email: chleecn@szu.edu.cn; lichaoh2@mail.sysu.edu.cn}

\affiliation{$^{1}$Guangdong Provincial Key Laboratory of Quantum Engineering and Quantum Metrology, School of Physics and Astronomy, Sun Yat-Sen University (Zhuhai Campus), Zhuhai 519082, China}
\affiliation{$^{2}$College of Physics and Optoelectronic Engineering, Shenzhen University, Shenzhen 518060, China}
\affiliation{$^{3}$State Key Laboratory of Optoelectronic Materials and Technologies, Sun Yat-Sen University (Guangzhou Campus), Guangzhou 510275, China}

\begin{abstract}
We uncover the interaction-induced \emph{stable self-localization} of bosons in disorder-free superlattices.
In these nonthermalized multi-particle states, one of the particles forms a superposition of multiple standing waves, so that it provides a quasirandom potential to localize the other particles.
We derive effective Hamiltonians for self-localized states and find their energy level spacings obeying the Poisson statistics for Anderson-like localization.
Surprisingly, we find that the correlated self-localization can be solely induced by interaction in the well-studied nonintegrable Bose-Hubbard models, which has been overlooked for a long time.
We propose a dynamical scheme to detect self-localization, where long-time quantum walks of a single particle form a superposition of multiple standing waves for trapping the subsequently loaded particles.
Our work provides an experimentally feasible way to realize stable Anderson-like localization in translation-invariant disorder-free systems.

\end{abstract}
\date{\today}
\maketitle

{\it Introduction}. Localization phenomena have attracted tremendous interests, ignited by Anderson localization in a disordered system~\cite{Anderson1958,physicstoday} and boosted by its generalization to many-body localization~\cite{Gornyi2005,MBL_BAA,Annelreview_MBL,rmpMBL}.
These disorder-induced localizations retain the memory of the initial state for a long time, which have been experimentally observed in various systems involving ultracold atoms~\cite{schreiber2015observation,choi2016exploring,bernien2017probing,lukin2019probing}, ions~\cite{smith2016many,zhang2017observation} and light fields~\cite{schwartz2007transport,segev2013anderson}.
As a paradigm of ergodicity breaking, localized states obey Poisson statistics and the entropy area law, distinguished from Wigner-Dyson statistics and the entropy volume law for thermalized states~\cite{Annelreview_MBL,rmpMBL}. 
However, disorder is not the only factor that induces localization.
A natural way to localize particles is to reduce tunneling, either by applying a tilting field~\cite{StarkMBL,StarkMBL2,Zhang2021,StarkMBLexperiment,Experiment2} or by designing a flat band~\cite{flatbandPRB,flatbandNJP}.

The search for disorder-free localization can date back to the seminal study in a mixture of $^3\rm{He}$ and $^4\rm{He}$ atoms~\cite{Kagan_1974,Kagan_1984}, where light particles are trapped by an effective quasistatic potential provided by heavy particles~\cite{lightheavyPRB,Annalphysics2015}, or vise versa in a qubit array coupled to waveguide~\cite{waveguideQED,ANPPRA}.
However, such an interaction-induced localization persists only for a short time~\cite{Quasi_localization2016,waveguideQED}.
A big step toward stable disorder-free localization is to use local constraints imposed by gauge symmetry~\cite{spinfermionPRL,spinfermionPRL2}, which turns out to be an extensive number of local conserved quantities that break ergodicity~\cite{gauge2018,Smith2018,quantumlinkPRL,Hart2021,quantumwalkPRA}.
Recently, lattice gauge theories have been simulated with ultracold atom systems~\cite{gorg2019realization,schweizer2019floquet,mil2020scalable,yang2020observation}, in particular, gauge invariance has been observed via ultracold Bose atoms in an optical superlattice~\cite{yang2020observation}.
However, the gauge-breaking errors therein will inevitably destroy disorder-free localization~\cite{Halimeh2022}.
In nonintegrable Bose-Hubbard models characterized by chaos and thermalization~\cite{kolovsky2004quantum,Sorg2014,islam2015measuring}, it is a highly nontrivial question whether a stable disorder-free localization exists without gauge symmetry.

\begin{figure}[!htp]
    \includegraphics[width=0.99\columnwidth]{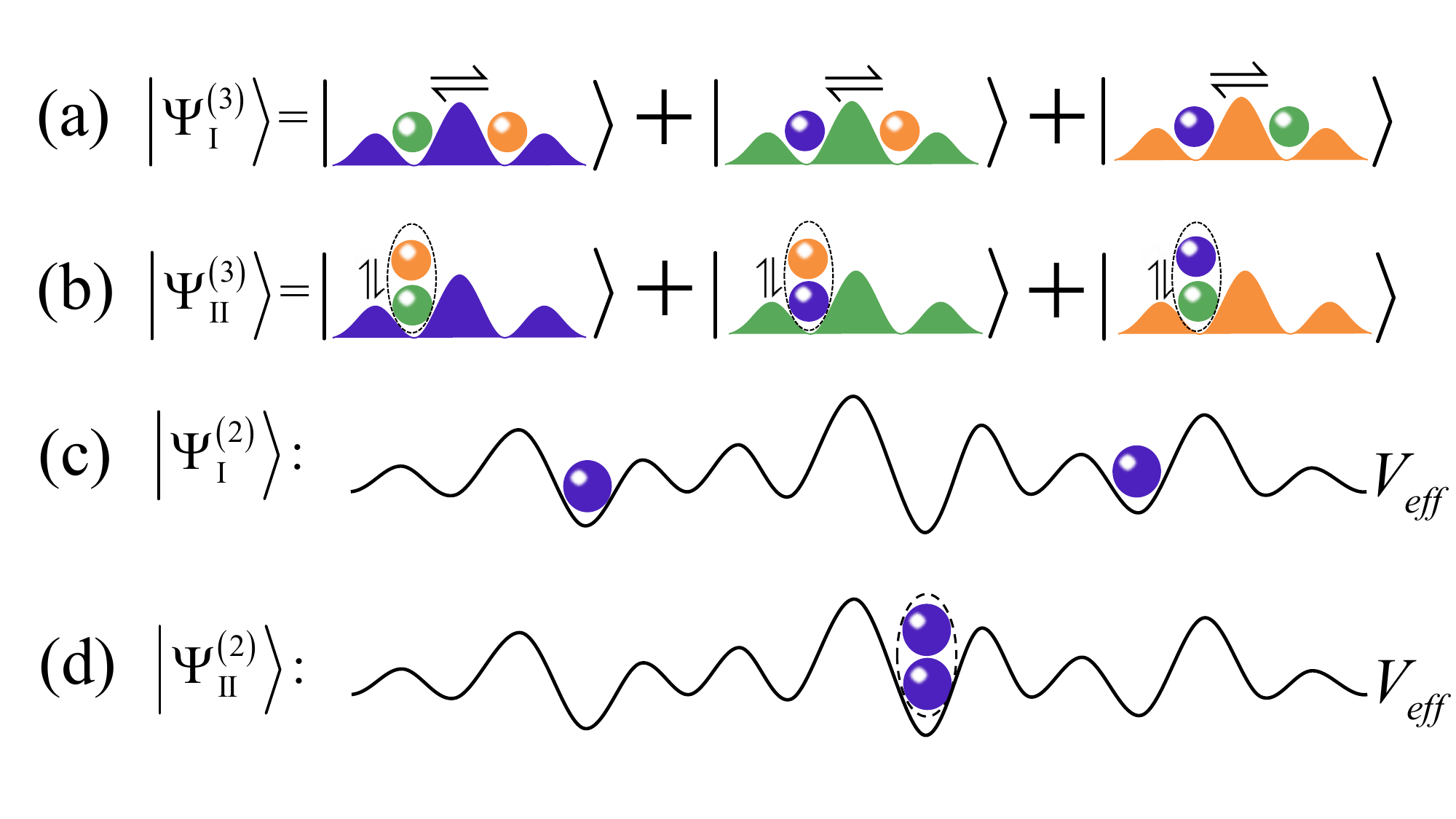}
   \caption{\label{Fig_sketch}Schematic illustration of (a) independent and (b) correlated Anderson-like localizations for three-boson states. The permutations between the particles compensate the indistinguishability of bosons.
   The effective potentials for (c) independent and (d) correlated Anderson-like localization, in which one particle provides the multi-standing-wave superposition for trapping the other two particles at different sites and the same site, respectively. 
  }
 \end{figure}

In this Letter, we uncover \emph{stable self-localization} of bosons in a disorder-free superlattice. 
As illustrated in Figs.~\ref{Fig_sketch}(a) and (b), there exist independent and correlated self-localizations, in which one of the particles forms the multi-standing-wave superposition and the other particles are trapped in different sites and the same site, respectively. 
We \emph{analytically} derive an effective Hamiltonian to explain the self-localization, that is, the superlattice and the multi-standing-wave superposition provide an irregular potential to trap the other particles [Figs.~\ref{Fig_sketch}(c) and (d)], which is dubbed Anderson-like localization (ALL).
The interaction-induced ALL is associated with the Poisson statistics of level spacings in the effective Hamiltonian, in stark contrast to the regular Landau levels in previous works~\cite{waveguideQED,poshakinskiy2021quantum}.
The Poisson statistics indicates unexpected emergent nonthermalized states in the nonintegrable Bose-Hubbard model which always leads to thermalization.
Remarkably, even for interacting bosons in a simple lattice, we find that correlated self-localization can appear under strong interaction, which has been overlooked in the long history of the well-studied Bose-Hubbard model. 
We show that one may employ long-time quantum walks of a single particle to form multi-standing-wave superposition, which traps the subsequently and adiabatically loaded particles.

\indent {\it Model and General Method}. We consider $N$ interacting bosons in a finite superlattice with $L$ sites, which obey the Bose-Hubbard type Hamiltonian,
\begin{equation}\label{Eq_Hami}
\hat H^{(N)}=-J\sum_{j=1}^{L-1}(\hat b_{j}^{\dagger}\hat b_{j+1}+h.c.)+ \sum_{j=1}^{L}\big[V_{j} \hat n_{j}+\frac{U}{2}\hat n_{j}\left( \hat n_{j}-1\right)\big].
\end{equation}
Here, $\hat{b}_{j}^{\dagger}$ ($\hat{b}_{j}$) is the bosonic creation (annihilation) operator at the $j$th site and $\hat n_{j}=\hat b_{j}^{\dagger}\hat b_{j}$ is the particle number operator. 
${V_{j}}={V\cos}\left[2\pi \beta \left(j+1/2\right)+\xi\right]$ is a spatially modulated potential with the modulation strength $V$, the modulation frequency $\beta=p/q$ ($p$ and $q$ are coprime integers), and the modulation phase $\xi$. 
When $\beta=1$, the superlattice becomes a simple lattice.  
$J$ and $U$ are strengths of the hopping and the on-site interaction, respectively.
In the following, we analyze the $N$-particle eigenstates under open boundary condition,
\begin{equation}\label{Eq_H}
	|\Psi^{(N)}\rangle=\sum_{i_1,i_2,...,i_N}\psi_{i_1,i_2,...,i_N}|i_1,i_2,...,i_N\rangle,
\end{equation}
where $\{i_1,i_2,...,i_N\}$ are positions of the $N$ bosons.
%
%
%
%
We find a missing jigsaw of eigenstates, 
%
\begin{eqnarray}\label{Eq_Anas}
\psi_{i_1,i_2,...,i_N}&\approx&\varphi_{i_1}\chi_{i_2,i_3,...,i_N}+\varphi_{i_2}\chi_{i_1,i_3,...,i_N}+ ... \nonumber \\
&+&\varphi_{i_{N-1}}\chi_{i_1,i_2,...,i_{N}}+\varphi_{i_N}\chi_{i_1,i_2,...,i_{N-1}},
\end{eqnarray}
where $\varphi$ is the extended single-particle wavefunction and $\chi$ is the localized $(N-1)$-particle wavefunction. We derive an effective Hamiltonian for the $(N-1)$ localized bosons. 
By substituting Eq.\eqref{Eq_Anas} into the Schr$\ddot{\rm{o}}$dinger equation $H\Psi=\epsilon \Psi$, and tracing the freedom of the single-particle state~\cite{poshakinskiy2021quantum}, we find that the localized $(N-1)$-particle states $\chi$ are approximately eigenstates of the effective Hamiltonian (see~\cite{SUP} for details),
\begin{eqnarray}\label{Eq_Eff}
\hat H_{eff}^{(N-1)}&=&\hat H^{(N-1)}+2U\sum_{j=1}^L |\varphi_j|^2\hat n_j. 
\end{eqnarray}
Here, the onsite potential is modified by the interaction and becomes quasirandom, $V_{eff}(j)=V_{j}+2U|\varphi_{j}|^2$.

\begin{figure}[!htp]
    \includegraphics[width=0.99\columnwidth]{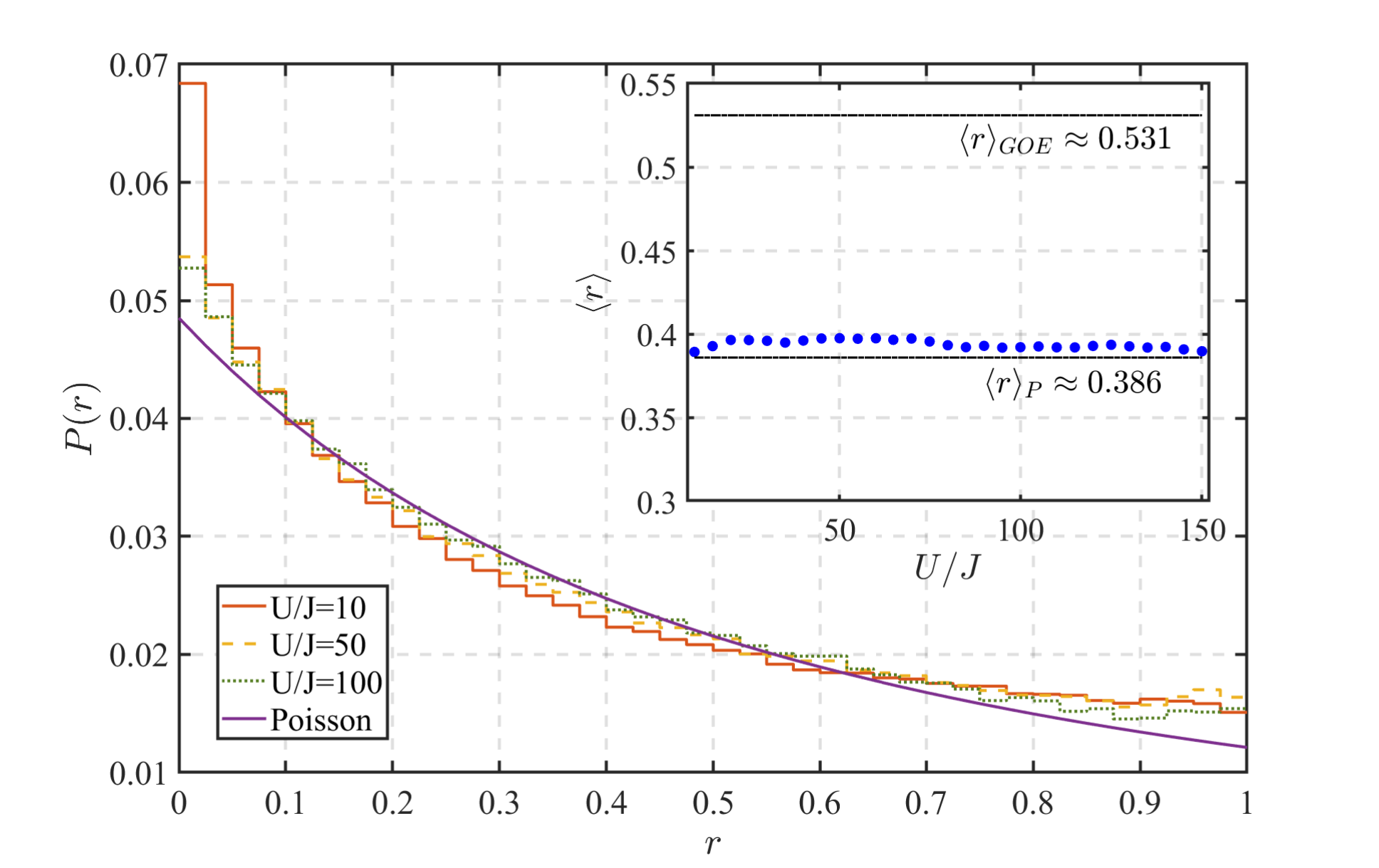}
   \caption{\label{Fig_energy_level_spacing}Energy level spacing distribution $P(r)$ of the effective Hamiltonians for different interaction strength $U/J$. The inset shows the average of the adjacent gap radio $\langle r \rangle$ for different $U/J$. The parameters are chosen as $L=64$, $N=2$, $p/q=1/4$, and $V/J=10$. The ranges of the modulated phase $\xi/2 \pi$ are chosen around $\xi_n\pm 0.002\pi$ with $\xi_n=\pi/4+n \pi/2$ ($n=0,1,2,3$). The sample interval of $\xi/2 \pi$ is $5 \times 10^{-5}$.
  }
\end{figure}

To explore the origin of self-localization, without loss of generality, we calculate the average ratio of the adjacent energy gap $\langle r \rangle$ and the level spacing statistics $P(r)$ for $H_{eff}^{(N-1)}$, where the ratio $r$ is defined as~\cite{Atas2013,Zhang2021}
\begin{equation}\label{Eq_energy_level}
r_m=\rm{min}(\delta_m, \delta_{m+1})/\rm{max}(\delta_m, \delta_{m+1}),
\end{equation}
with $\delta_m=E_m-E_{m+1}$, the energy gap between the eigenvalues $E_{m}$ and $E_{m+1}$ of $H_{eff}^{(N-1)}$. 
In two-particle system, most self-localized states appear around $\xi_n=\pi/4+n \pi/2$ ($n=0,1,2,3$).
The conditions to distinguish self-localized states are given in the Supplemental Materials~\cite{SUP}.
By choosing the modulation phase around $\xi_n\pm 0.002\pi$, for each self-localized state, we can decompose the state and acquire $\varphi_j$ for an effective Hamiltonian which is a sample used for level spacing statistics. 
The energy level spacing distributions for different interactions ($U/J=10,\ 50,\ 100$) are shown in Fig.~\ref{Fig_energy_level_spacing}.
The other parameters are chosen as $L=64$, $N=2$, $p/q=1/4$ and $V/J=10$.
We find that the level spacing statistics $P(r)$ are close to the Possion statistics (the purple solid line).
Furthermore, we show the average ratio $\langle r\rangle$ versus the interaction strength in the inset of Fig.~\ref{Fig_energy_level_spacing}, which is around $\langle r\rangle=0.386$, the value predicted for the Possion statistics~\cite{Atas2013,Zhang2021}. 
The level spacing statistics indicate that the particle is localized in the effective quasirandom potentials induced by the interaction (see~\cite{SUP} for details).
Therefore, we call this exotic localization the interaction-induced ALL.

\begin{figure*}[!htp]
	\includegraphics[width=1\linewidth]{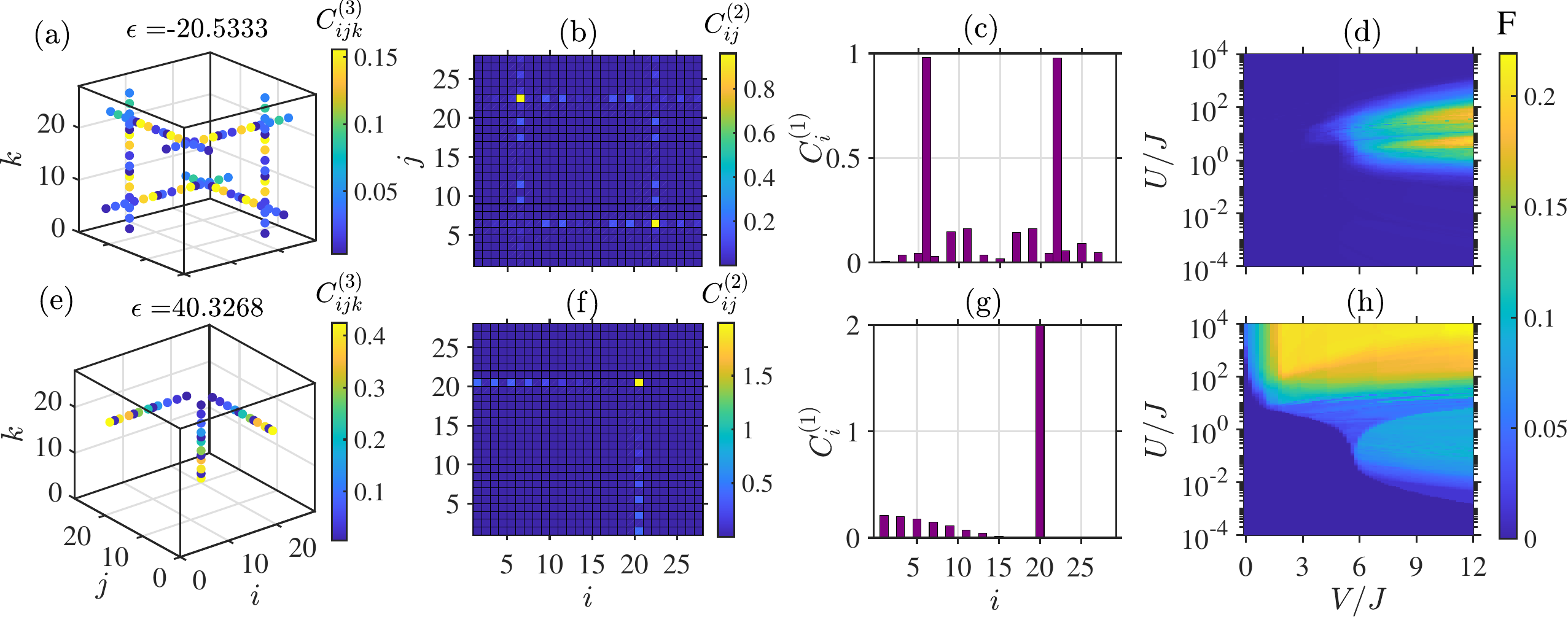}
	\caption{\label{Fig_Three1}(a)-(c) The third-, second- and first-order correlation functions for the independent ALL respectively. The first-order correlation function is also called density distribution. (d) Fraction of the number of independent self-localized states in total eigenstates as a function of interaction strength ($U/J$) and the modulation strength of the onsite potential ($V/J$). (e)-(h) The same as (a)-(d), but for correlated ALL. The system parameter are set as $p/q=1/4$, $V/J=10$, $U/J=20$, $\xi=-\beta\pi$, $L=28$ for (a)-(c) and (e)-(g), $L=16$ for (d) and (h).
}\end{figure*}

{\it Three-particle Self-localized States.} 
Below we illustrate the interaction-induced ALL in three-particle systems.
By solving the equation $\hat{H}^{(3)}\Psi=\epsilon\Psi$, we acquire three-particle states $|\Psi^{(3)}\rangle=\sum_{i_1,i_2,i_3}\psi_{i_1,i_2,i_3}|i_1,i_2,i_3\rangle$ with $\psi_{i_1,i_2,i_3}=\psi_{i_1,i_3,i_2}=\psi_{i_3,i_1,i_2}=\psi_{i_3,i_2,i_1}=\psi_{i_2,i_3,i_1}=\psi_{i_2,i_1,i_3}$. 
To show their spatial distributions, we analyze their correlation functions,
 \begin{eqnarray} C_{i}^{(1)}&=&\langle\Psi^{(3)}|\hat{a}_i^{\dagger}\hat{a}_i|\Psi^{(3)}\rangle,\nonumber \\
 C_{ij}^{(2)}&=&\langle\Psi^{(3)}|\hat{a}_i^{\dagger}\hat{a}_j^{\dagger}\hat{a}_j\hat{a}_i|\Psi^{(3)}\rangle,\nonumber \\
C_{ijk}^{(3)}&=&\langle\Psi^{(3)}|\hat{a}_i^{\dagger}\hat{a}_j^{\dagger} \hat{a}_k^{\dagger}\hat{a}_k\hat{a}_j\hat{a}_i|\Psi^{(3)}\rangle.
\end{eqnarray}
$C_{i}^{(1)}, C_{ij}^{(2)}, C_{ijk}^{(3)}$ are the first-, second-, and third-order correlation functions for $\Psi^{(3)}$, respectively. 
We calculate the correlation functions for two typical self-localized states with energies $\epsilon=-20.5333$ and $\epsilon=40.3268$; see Figs.~\ref{Fig_Three1}(a)-(c) and (e)-(g).
The parameters are chosen as $V=10J$, $U=20J$, $\beta=p/q=1/4$, $\xi=-\beta\pi$ and $L=28$.
For both cases, one of the three bosons has a broad spatial distribution.
However, the other two bosons are localized at different sites (the $6$th and $22$th sites) and the same site (the $20$th site) for $\epsilon=-20.5333$ and $\epsilon=40.3268$, respectively. 
Therefore, we classify the localization phenomenon into (i) independent ALL where the two bosons locate at two different sites and (ii) correlated ALL where the two bosons locate at the one site (i.e. form an on-site bound pair). 
The two particles can randomly localize at some lattice sites for the independent ALL, while they can localize at any lattice site for the correlated ALL (see~\cite{SUP} for details).

We show how to numerically decompose the above self-localized states into the form of Eq.~\eqref{Eq_Anas}. 
$\psi_{i_1,i_2,i_3}$ is an element of a $(L \times L \times L)$ tensor, which can be reshaped into a $(L \times L ^2)$ matrix with elements $\tilde{\psi}_{i_1,r}$ [$r=\left(i_2-1\right)\times L+i_3$]. 
We perform singular value decomposition (SVD) on the states, $\tilde{\psi}_{i_{1},r}=\sum_{m,n}S_{i_{1},m}D_{m,n}W_{n,r}$. 
For independent ALLs, we find that the singular values are dominated by three terms and their values are quite close, that is, $D_{11}\approx D_{22}\approx D_{33}$.
Furthermore, a common extended single-particle state $\varphi$ is close to three single-particle standing waves $\{{s}^{(f)}\approx\mu^{(f)}\approx\nu^{(f)}\}$ hidden in the matrices of $S$ and $W$~\cite{SUP}.
Both $\varphi$ and $\chi$ can be extracted by further SVD operation of $S$ and $W$.
For correlated ALLs, the singular values are dominated by two terms $D_{11}$ and $D_{22}$, and a common $\varphi$ is obtained via two single-particle standing waves ($\{{s}^{(f)}\approx\mu^{(f)}\}$) (see~\cite{SUP} for more details).

Once $\varphi$ is obtained, by diagonalizing the Hamiltonian Eq.~\eqref{Eq_Eff} we can alternatively obtain the two-particle localized states $\chi'$, and then give the three-particle state ${\Psi}'^{(3)}$ according to Eq.~\eqref{Eq_Anas}.
The $\chi'$ with the highest fidelity $|\langle \Psi^{(3)}|{\Psi'}^{(3)}\rangle|$ is the desired two-particle localized state. 
For the self-localized states shown in Figs.~\ref{Fig_Three1}(a) and \ref{Fig_Three1}(e), the maximum fidelities are quite close to 1. 
%
%
Note that a large group of localized states can be obtained with the effective Hamiltonian, leaving alone the two states mentioned above.

The independent and correlated ALLs are quite generic in Bose-Hubbard systems. 
Numerically, we successfully single out independent (correlated) self-localized states that satisfy: (i) the sum of the first three (two) singular values is greater than 0.8; (ii) the fidelities among the extended single-particle states $\{{s}^{(f)},\mu^{(f)},\nu^{(f)}\}$ ($\{{s}^{(f)},\mu^{(f)}\}$) are greater than 0.9; (iii) the inverse participation radio (IPR) of $\chi$ is greater than $0.4$ (0.8), where IPR=$\sum_{i,j}|\chi_{i,j}|^4/[\sum_{i,j}|\chi_{i,j}|^2]^2$. 
By varying the interaction ($U$) and modulation strength of the onsite potential ($V$), we show the fraction of the two types of self-localized states over all eigenstates in Figs.~\ref{Fig_Three1} (d) and (h), respectively. 
Both independent and correlated self-localized states disappear when $U=0$.
The independent ALL appears only when the spatial modulation of the onsite potential is strong enough and the interaction strength is modest.
It means that the independent ALL results from the interplay between interaction and spatial modulation of the onsite potential.
More explicitly, spatial modulation of the onsite potential leads to flat bands which facilitate localization~\cite{SUP}.
However, correlated ALLs still exist in the absence of the modulated on-site potential, that is, they are purely induced by the strong interaction.
Due to the strong interaction, the correlated bound states have a much smaller group velocity, which can be easily captured by the background potential provided by the third particle.

Apart from independent and correlated ALLs, there exists another kind of self-localization, i.e., one of the three particles is localized, whereas the other two particles are extended, which cannot be captured by our effective Hamiltonian. 
 In such states, the third-order correlation functions behave as three intersecting planes. 
 Like the independent ALL, the localized particle can localize randomly at some of the sites.
 Furthermore, such localized states also exist when $V$ is greater than a certain value and under moderate interaction strength (see~\cite{SUP} for details).

\begin{figure}[!]
  \includegraphics[width=1.1\linewidth]{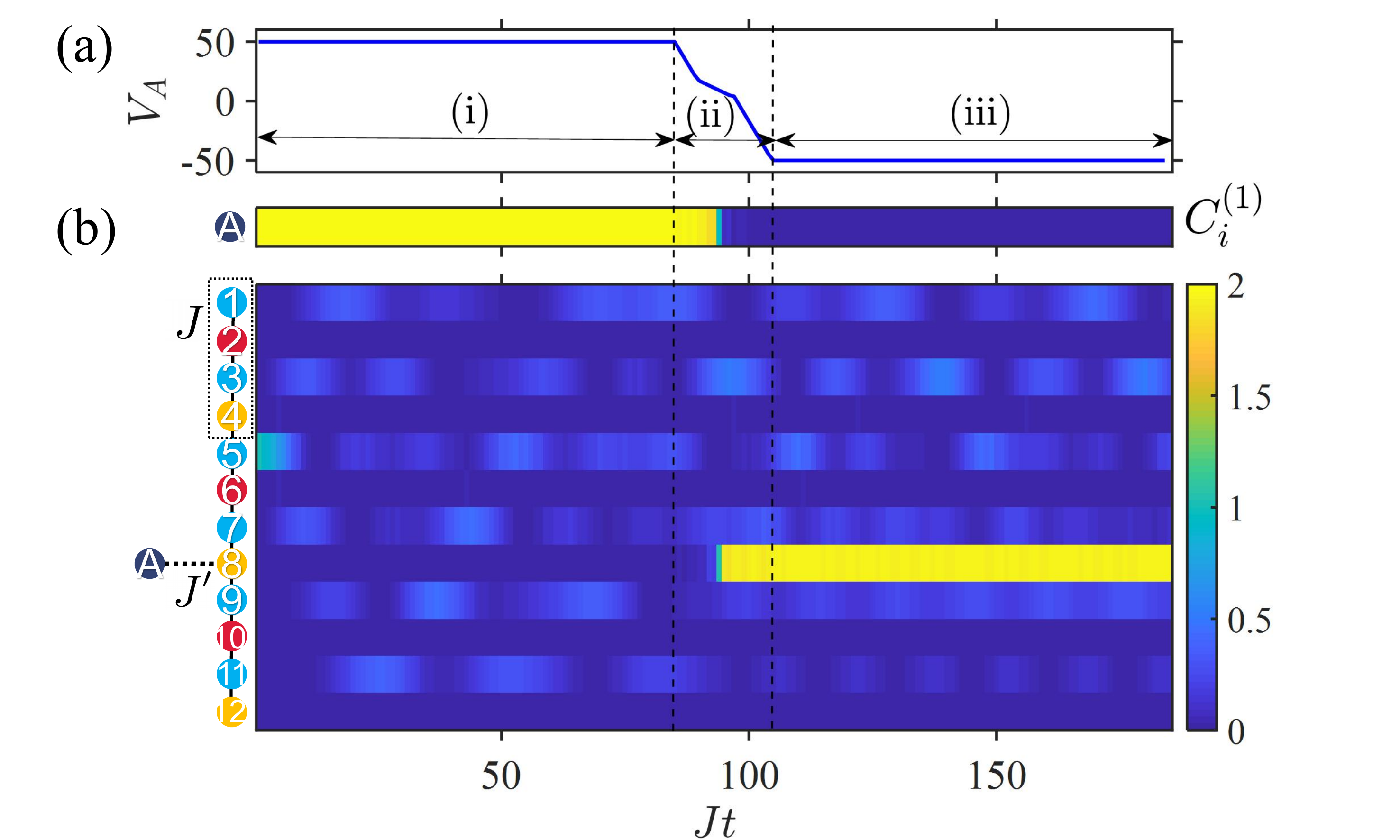}
  \caption{\label{Fig_dy}(a) The varying potential $V_A$ of the auxiliary site $A$ versus time during three steps.  
   (b) The time evolution of density distribution for the correlated ALL. The hopping amplitude between the auxiliary site $A$ and superlattice are chosen as $J'=0$ in processes (i) and (iii),  and $J'=4J$ in process (ii). The other parameters are set as $T_{1}=84J/\hbar$, $T_{2}=104J/\hbar$, $T_{3}=184J/\hbar$, $L=12$, $U=20J$, {$V=10J$, $\xi=-\beta\pi$}, and $p/q=1/4$. 
  }
\end{figure}

 {\it Dynamical Simulation Scheme}. To observe ALLs, a key problem is to create the effective potential $V_{eff}$ formed by standing waves, which is an equal-weight superposition of Bloch states with opposite momenta (see~\cite{SUP} for details). 
 Such states can be prepared by single-particle quantum walks in a finite superlattice reflected by the boundaries repeatedly.
 The Bloch states with opposite momenta spread throughout the superlattice and form a standing wave in the long-time evolution. 
Then we adiabatically load the other two bosons into the lattice.
Once the loading is complete, the two bosons can be stably localized in the effective potential.

The scheme consists of three steps: (i) $0<t<T_{1}$, creation of the effective potential; (ii) $T_1<t<T_2$, loading particles into the effective potential; and (iii) $T_2<t<T_3$, free evolution of ALLs.
We take the correlated ALL as an example to illustrate the progress (similarly, see~\cite{SUP} for the independent ALL).
In process (i), we prepare one boson at the $5$th site of the superlattice with 3 unit cells containing $12$ sites ($p/q=1/4$). 
While the boson in the superlattice undergoes quantum walks for a long time $T_1=84J/\hbar$, the other two bosons are trapped at the auxiliary site $A$, which is isolated from the superlattice.
This can be achieved by setting a large potential bias and zero hopping strength $J'=0$ between the auxiliary site and the superlattice.
In process (ii), we gradually reduce the potential bias and turn on the hopping $J'=4J$, so that the two trapped particles will be adiabatically and completely transferred to the superlattice at the time $T_2=104J/\hbar$. 
In process (iii), we turn off the hopping $J'=0$, and let the three-particle states undergo a free evolution to time $T_3=184J/\hbar$. 
The potential of the auxiliary site $V_A$ versus time during the three steps is schematically shown in Fig.~\ref{Fig_dy}(a), in which the varying rates of potential in process (ii) change three times for efficient transfer (see ~\cite{SUP} for details).  

The density evolution of the three-particle state is shown in Fig.~\ref{Fig_dy}(b).
In the long period of process (iii), the two bosons are localized at the site where they are loaded, confirming successful observations of the correlated ALLs.
The success depends on a large proportion of self-localized states in the three-particle state $\Psi(T_{2})$ at time $T_2$.
We project $\Psi(T_{2})$ onto eigenstates of the Hamiltonian at the beginning and find that the proportion of self-localized states is about $94\%$.
The localization diffuses after a long-time evolution without interaction  ($U=0$)~\cite{SUP}, indicating that ALL is indeed induced by the interaction in the superlattice.

{\it Conclusion and Discussion.\label{Sec5}}
We have revealed interaction-induced partial self-localization in disorder-free Bose-Hubbard systems via a general semi-analytical method.
The extended single-particle wavefunction provides an effective potential for trapping the other particles, and the depth of the effective potential depends on the interaction strength. 
Importantly, the correlated Anderson-like localization, which may exist in the absence of on-site-potential modulation, is purely induced by interaction.
We also propose a dynamical simulation scheme to observe the Anderson-like localization.

Theoretically, our work will motivate more fundamental research on the emergent phenomena induced by interaction.  
The deep relation between the interaction-induced Anderson-like localization and lattice gauge theories needs to be clarified in further study~\cite{yang2020observation}.
Stable self-localized states open an avenue to explore novel nonthermalized states embedded in a thermalizing spectrum, such as many-body scars~\cite{turner2018weak}.
Experimentally, Anderson-like localization persists in small-size ($L\sim10$) systems, which are readily accessible with superconducting qubits ~\cite{superconductingPRL} and ultracold atoms~\cite{islam2015measuring,yang2020observation} etc.

\begin{acknowledgments}{We acknowledge Xiaoying Du and Wenjie Liu for useful discussions. This work is supported by the NSFC (Grants No. 12025509, No. 11874434), the Key-Area Research and Development Program of GuangDong Province (Grants No. 2019B030330001), and the Science and Technology Program of Guangzhou (China) (Grants No. 201904020024). Y.K. is partially supported by the National Natural Science Foundation of China (Grant No. 11904419).  Li Zhang is partially supported by the Fundamental Research Funds for the Central Universities, Sun Yat-sen University (Grant No. 22qntd3101).}
\end{acknowledgments}

\bibliography{main}

\onecolumngrid
\clearpage

\renewcommand {\Im}{\mathop\mathrm{Im}\nolimits}
\renewcommand {\Re}{\mathop\mathrm{Re}\nolimits}
\renewcommand {\i}{{\rm i}}
\renewcommand {\phi}{{\varphi}}

\begin{center}
	\noindent\textbf{\large{Supplemental Material:}}
	\\\bigskip
	\noindent\textbf{\large{Stable interaction-induced Anderson-like localization embedded in standing waves}}
	\\\bigskip
	\onecolumngrid
	
	Na Zhang$^{1,2,3}$, Yongguan Ke$^{1,2,*}$, Ling Lin$^{2}$, Li Zhang$^{2}$, Chaohong Lee$^{1,2,3,\dag}$
	
	\small{$^1$ \emph{Guangdong Provincial Key Laboratory of Quantum Engineering and Quantum Metrology, School of Physics and Astronomy, Sun Yat-Sen University (Zhuhai Campus), Zhuhai 519082, China}}\\
	\small{$^2$ \emph{College of Physics and Optoelectronic Engineering, Shenzhen University, Shenzhen 518060, China}}\\
	\small{$^3$ \emph{State Key Laboratory of Optoelectronic Materials and Technologies, Sun Yat-Sen University (Guangzhou Campus), Guangzhou 510275, China}}\\
\end{center}

\setcounter{equation}{0}
\setcounter{table}{0}
\renewcommand{\theequation}{S\arabic{equation}}

\renewcommand{\thesection}{S\arabic{section}}

\setcounter{equation}{0}
\newcounter{sfigure}
\setcounter{sfigure}{1}
\setcounter{table}{0}

\renewcommand{\theequation}{S\arabic{equation}}

 \renewcommand\thefigure{S{\arabic{figure}}}
\renewcommand{\thesection}{S\arabic{section}}



\section{S1. Derivation of the general method}
 \noindent In this section, from the $N$-particle Schr$\ddot{\rm{o}}$dinger equation, we intend to derive the effective Hamiltonian for the $(N-1)$ particles localized by one of the $N$ particles. The Hamiltonian  can be written as
\begin{equation}
H^{(N)}=H_{0}+U,
\end{equation}
where the one-body Hamiltonian
\begin{gather}
H_{0}=H^{1}\otimes I^{2}...\otimes I^{N}+I^{1}\otimes H^{2}...\otimes I^{N}+...+I^{1}\otimes I^{2}...\otimes H^{N},
\end{gather}
and the two-body interacting Hamiltonian
\begin{equation}
U_{i_1,i_2,i_3...i_N}^{j_1,j_2,j_3...j_N}=U\delta_{i_1,j_1}\delta_{i_2,j_2}\delta_{i_3,j_3}...\delta_{i_N,j_N}
\left(\delta_{i_1,i_2}+\delta_{i_1,i_3}+...+\delta_{i_1,i_N}+\delta_{i_2,i_3}...+\delta_{i_{N-1},i_N}\right).
\end{equation}
The eigenstates are given by
\begin{equation}
|\Psi\rangle=\sum_{i_1,i_2...i_N}\psi_{i_1,i_2...i_N}|i_1,i_2...i_N\rangle.
\end{equation}
The eigenstates of 1 extended particle and N-1 localized particles can be well described by the following ansatz,
\begin{equation}\label{Eq_ansatz}
\psi_{i_1,i_2...i_N}\approx\varphi^{l}_{i_1}\chi_{i_2,i_3...i_N}+\varphi^{l}_{i_2}\chi_{i_1,i_3...i_N}+...+\varphi^{l}_{i_N}\chi_{i_1,i_2...i_N-1},
\end{equation}
where $\varphi$ is the wavefunction of the extended particle, and $\chi$ is the wavefunction of the $N-1$ localized particles. Our following task is to obtain the localized state $\chi$ for $N-1$ particles by solving the many-body Schr$\ddot{\rm{o}}$dinger equation,

\begin{equation}
\begin{split}
\sum_{i_1,i_2...i_N}\left(H^{1}\otimes I^{2}...\otimes I^{N}+I^{1}\otimes H^{2}...\otimes I^{N}+...+I^{1}\otimes I^{2}...\otimes H^{N}\right)_{i_1,i_2...i_N}^{j_1,j_2...j_N}\psi_{i_1,i_2...i_N}\\
+\sum_{i_1,i_2...i_N}U_{i_1,i_2...i_N}^{j_1,j_2...j_N}=E\psi_{j_1,j_2...j_N}.
\end{split}	
\end{equation}
Substituting the ansatz Eq.\eqref{Eq_ansatz} into the Schr$\ddot{\rm{o}}$dinger equation, we have

 \begin{equation}\label{Eq_7}
		\begin{split}
    &	\sum_{i_1,i_2...i_N}\left(H^{1}\otimes I^{2}...\otimes I^{N}\right)_{i_1,i_2...i_N}^{j_1,j_2...j_N}\left(\varphi_{i_1}^{l}\chi_{i_2,i_3...i_N}+\varphi^{l}_{i_2}\chi_{i_1,i_3...i_N}+...+\varphi^{l}_{i_N}\chi_{i_1,i_2...i_N-1}\right)\\
    & +\sum_{i_1,i_2...i_N}\left(I^{1}\otimes H^{2}...\otimes
     I^{N}\right)_{i_1,i_2...i_N}^{j_1,j_2...j_N}\left(\varphi^{l}_{i_1}
       \chi_{i_2,i_3...i_N}+\varphi^{l}_{i_2}\chi_{i_1,i_3...i_N}+...+\varphi^{l}_{i_N}\chi_{i_1,i_2...i_N-1}\right)+...\\
    &+\sum_{i_1,i_2...i_N}\left(I^{1}\otimes I^{2}...\otimes H^{N}\right)_{i_1,i_2...i_N}^{j_1,j_2...j_N}\left(\varphi^{l}_{i_1}\chi_{i_2,i_3...i_N}+\varphi^{l}_{i_2}\chi_{i_1,i_3...i_N}+...+\varphi^{l}_{i_N}\chi_{i_1,i_2...i_N-1}\right)\\
    &+\sum_{i_1,i_2...i_N}U_{i_1,i_2...i_N}^{j_1,j_2...j_N}\left(\varphi^{l}_{i_1}\chi_{i_2,i_3...i_N}+\varphi^{l}_{i_2}\chi_{i_1,i_3...i_N}+...+\varphi^{l}_{i_N}\chi_{i_1,i_2...i_N-1}\right)\\
    & =E\left(\varphi^{l}_{j_1}\chi_{j_2,j_3...j_N}+\varphi^{l}_{j_2}\chi_{j_1,j_3...j_N}
     +...+\varphi^{l}_{j_N}\chi_{j_1,j_2...j_N-1}\right).\\
    		\end{split}	
\end{equation}
We assume that $\varphi$ is approximately the eigenstate of the single-particle Hamiltonian, i.e., $H^{(1)}\varphi^{(l)}\approx\epsilon_l\varphi^{(l)}$. For the sake of brevity, we omit the summation notation and simplify $\varphi^{l}$ as $\varphi$; Eq.\eqref{Eq_7} becomes

\begin{equation}
		\begin{split}
    & \epsilon_{l}\varphi_{j_1}\chi_{j_2,j_3...j_N}+H^{1}_{i_1,j_1}\left(\varphi_{j_2}\chi_{i_1,j_3...j_N}+...
    +\varphi_{j_N}\chi_{i_1,j_2...j_N-1}\right) \\
    &  +\epsilon_{l}\varphi_{j_2}\chi_{j_1,j_3...j_N}+H^{2}_{i_2,j_2}\left(\varphi_{j_1}\chi_{i_2,j_3...j_N}...+\varphi_{j_N}\chi_{j_1,i_2...j_N-1}\right) +...\\
    & +\epsilon_{l}\varphi_{j_N}\chi_{j_1,j_2...j_N-1}+H^{N}_{i_N,j_N}\left(\varphi_{j_1}\chi_{j_2,j_3...i_N}+\varphi_{j_2}\chi_{j_1,j_3...i_N}+...
    +\varphi_{j_N-1}\chi_{j_1,j_2...i_N}\right)\\
    & +U\left(\delta_{j_1,j_2}+\delta_{j_1,j_3}+...+\delta_{j_1,j_N}\right) \left(\varphi_{j_1}\chi_{j_2,j_3...j_N}+\varphi_{j_2}\chi_{j_1,j_3...j_N}+...+\varphi_{j_N}\chi_{j_1,j_2...j_N-1}\right)\\
    & +U\left(\delta_{j_2,j_3}+\delta_{j_2,j_4}+...+\delta_{j_2,j_N}\right) \left(\varphi_{j_1}\chi_{j_2,j_3...j_N}+\varphi_{j_2}\chi_{j_1,j_3...j_N}+...+\varphi_{j_N}\chi_{j_1,j_2...j_N-1}\right)\\
    & +U\left(\delta_{j_3,j_4}+\delta_{j_3,j_5}+...+\delta_{j_3,j_N}\right) \left(\varphi_{j_1}\chi_{j_2,j_3...j_N}+\varphi_{j_2}\chi_{j_1,j_3...j_N}+...+\varphi_{j_N}\chi_{j_1,j_2...j_N-1}\right)+...\\
    & +U\delta_{j_{N-1},j_N}
    \left(\varphi_{j_1}\chi_{j_2,j_3...j_N}+\varphi_{j_2}\chi_{j_1,j_3...j_N}+...+\varphi_{j_N}\chi_{j_1,j_2...j_N-1}\right)\\
    & =E\left(\varphi_{j_1}\chi_{j_2,j_3...j_N}+\varphi_{j_2}\chi_{j_1,j_3...j_N}
     +...+\varphi_{j_N}\chi_{j_1,j_2...j_N-1}\right).
    		\end{split}	
\end{equation}
Multiplying the equation by $\varphi_{j_1}^{*}$ and summing over $j_1$, the summation notation is omitted for brevity. 
And then we have
\begin{equation}\label{Eq_9}
		\begin{split}
    & \epsilon_{l}\chi_{j_2,j_3...j_N}+\epsilon_{l}\varphi_{i_1}^{*}\left(\varphi_{j_2}\chi_{i_1,j_3...j_N}+...
    +\varphi_{j_N}\chi_{i_1,j_2...j_N-1}\right) \\
    &  +\epsilon_{l}\varphi_{j_1}^{*}\varphi_{j_2}\chi_{j_1,j_3...j_N}+H^{2}_{i_2,j_2}\varphi_{j_1}^{*}\left(\varphi_{j_1}\chi_{i_2,j_3...j_N}...+\varphi_{j_N}\chi_{j_1,i_2...j_N-1}\right) +...\\
    & +\epsilon_{l}\varphi_{j_1}^{*}\varphi_{j_N}\chi_{j_1,j_2...j_N-1}+H^{N}_{i_N,j_N}\varphi_{j_1}^{*}\left(\varphi_{j_1}\chi_{j_2,j_3...i_N}+\varphi_{j_2}\chi_{j_1,j_3...i_N}+...
    +\varphi_{j_N-1}\chi_{j_1,j_2...i_N}\right)\\
    & +\delta_{j_1,j_2} U\left(|\varphi_{j_1}|^{2}\chi_{j_2,j_3...j_N}+\varphi_{j_1}^{*}\varphi_{j_2}\chi_{j_1,j_3...j_N}+...+\varphi_{j_1}^{*}\varphi_{j_N}\chi_{j_1,j_2...j_N-1}\right)\\
    & +\delta_{j_1,j_3} U\left(|\varphi_{j_1}|^{2}\chi_{j_2,j_3...j_N}+\varphi_{j_1}^{*}\varphi_{j_2}\chi_{j_1,j_3...j_N}+...+\varphi_{j_1}^{*}\varphi_{j_N}\chi_{j_1,j_2...j_N-1}\right)+...\\
    & +\delta_{j_1,j_N} U\left(|\varphi_{j_1}|^{2}\chi_{j_2,j_3...j_N}+\varphi_{j_1}^{*}\varphi_{j_2}\chi_{j_1,j_3...j_N}+...+\varphi_{j_1}^{*}\varphi_{j_N}\chi_{j_1,j_2...j_N-1}\right)\\
    & +\delta_{j_2,j_3} U\left(|\varphi_{j_1}|^2\chi_{j_2,j_3...j_N}+\varphi_{j_1}^{*}\varphi_{j_2}\chi_{j_1,j_3...j_N}+...+\varphi_{j_1}^{*}\varphi_{j_N}\chi_{j_1,j_2...j_N-1}\right)\\
    & +\delta_{j_2,j_4} U\left(|\varphi_{j_1}|^2\chi_{j_2,j_3...j_N}+\varphi_{j_1}^{*}\varphi_{j_2}\chi_{j_1,j_3...j_N}+...+\varphi_{j_1}^{*}\varphi_{j_N}\chi_{j_1,j_2...j_N-1}\right)+...\\
    & +\delta_{j_{N-1},j_N}U
    \left(|\varphi_{j_1}|^2\chi_{j_2,j_3...j_N}+\varphi_{j_1}^{*}\varphi_{j_2}\chi_{j_1,j_3...j_N}+...+\varphi_{j_1}^{*}\varphi_{j_N}\chi_{j_1,j_2...j_N-1}\right)\\
    & =E\left(\chi_{j_2,j_3...j_N}+\varphi_{j_1}^{*}\varphi_{j_2}\chi_{j_1,j_3...j_N}
     +...
    +\varphi_{j_1}^{*}\varphi_{j_N}\chi_{j_1,j_2...j_N-1}\right).
    		\end{split}	
\end{equation}
 We focus on $\chi$ which is strongly localized. We assume that $\chi$ is orthogonal to $\varphi$, and thus the terms $\propto \varphi^{*}_{n}\chi_{m_1,m_2,m_3...m_{N-1}}$ can be omitted if one of the indexes for $\chi$ is the same as that for $\varphi$, e.g., the term $\propto \sum_{j_1}\varphi^{*}_{j_1}\chi_{j_1,j_3,j_4...j_{N}}$. Then Eq.\eqref{Eq_9} can be simplified as
\begin{equation}
		\begin{split}
        & \epsilon\chi_{j_2,j_3...j_N}+H^{2}_{i_2,j_2}\chi_{i_2,j_3...j_N}+H^{3}_{i_3,j_3}\chi_{j_2,i_3...j_N}+...+H^{N}_{i_N,j_N}\chi_{j_2,j_3...i_N}
         \\
        & +U\left(2|\varphi_{j_2}|^{2}\chi_{j_2,j_3...j_N}+\varphi_{j_2}^{*}\varphi_{j_3}\chi_{j_2,j_2...j_N}...+\varphi_{j_2}^{*}\varphi_{j_N}\chi_{j_2,j_2...j_N-1}\right)\\
        & +U\left(2|\varphi_{j_3}|^{2}\chi_{j_2,j_3...j_N}+\varphi_{j_3}^{*}\varphi_{j_2}\chi_{j_3,j_3...j_N}+...+\varphi_{j_3}^{*}\varphi_{j_N}\chi_{j_3,j_2,j_3...j_N-1}\right)+...\\
        & +U\left(2|\varphi_{j_N}|^{2}\chi_{j_2,j_3...j_N}+\varphi_{j_N}^{*}\varphi_{j_2}\chi_{j_N,j_3...j_N}+...+\varphi_{j_N}^{*}\varphi_{j_N-1}\chi_{j_N,j_2...j_N-1}\right)\\
         &+U\left(\delta_{j_2,j_3}+\delta_{j_2,j_4}
          +...
         +\delta_{j_{N-1},j_N}\right)\chi_{j_2,j_3...j_N}\\
         &=E\chi_{j_2,j_3...j_N}.
    	\end{split}	
\end{equation}
From the equation above, we know that $\chi$ is the eigenstate of the following effective Hamiltonian,
\begin{equation}\label{Eq_Hamlast}
H_{eff}^{N-1}=-J\sum_{\langle ij\rangle}\hat b_{j}^{\dagger}\hat b_{j+1}+ \sum_{j}\left(V_{j}+2U|\varphi_{j}|^2\right)\hat{n_{j}}+\frac{U}{2}\sum_{j}\hat n_{j}\left( \hat n_{j}-1\right)+U\sum_{i,j}\left(\varphi^{*}_i\varphi_j\hat b_{i}^{\dagger}\hat b_{j}^{\dagger}\hat b_{j}\hat b_{j}+h.c.\right).
\end{equation}
The localization is induced by the effective potential $V_{eff}(j)=V_{j}+2U|\varphi_{j}|^2$. 
The last term is a hopping term, which intend to divide one particle apart from a bound state. 
The numerical calculations show that the effect of the last term can be safely neglected.
Therefore, we can obtain the effective Hamiltonian for $(N-1)$ self-localized particles, 
\begin{equation}\label{Eq_Hamlast}
H_{eff}^{(N-1)}=-J\sum_{\langle ij\rangle}\hat b_{j}^{\dagger}\hat b_{j+1}+ \sum_{j}\left(V_{j}+2U|\varphi_{j}|^2\right)\hat{n_{j}}+\frac{U}{2}\sum_{j}\hat n_{j}\left( \hat n_{j}-1\right).
\end{equation}

\section{S2. Two-particle system}
\begin{figure}[!htp]
  \includegraphics[width=0.8\columnwidth]{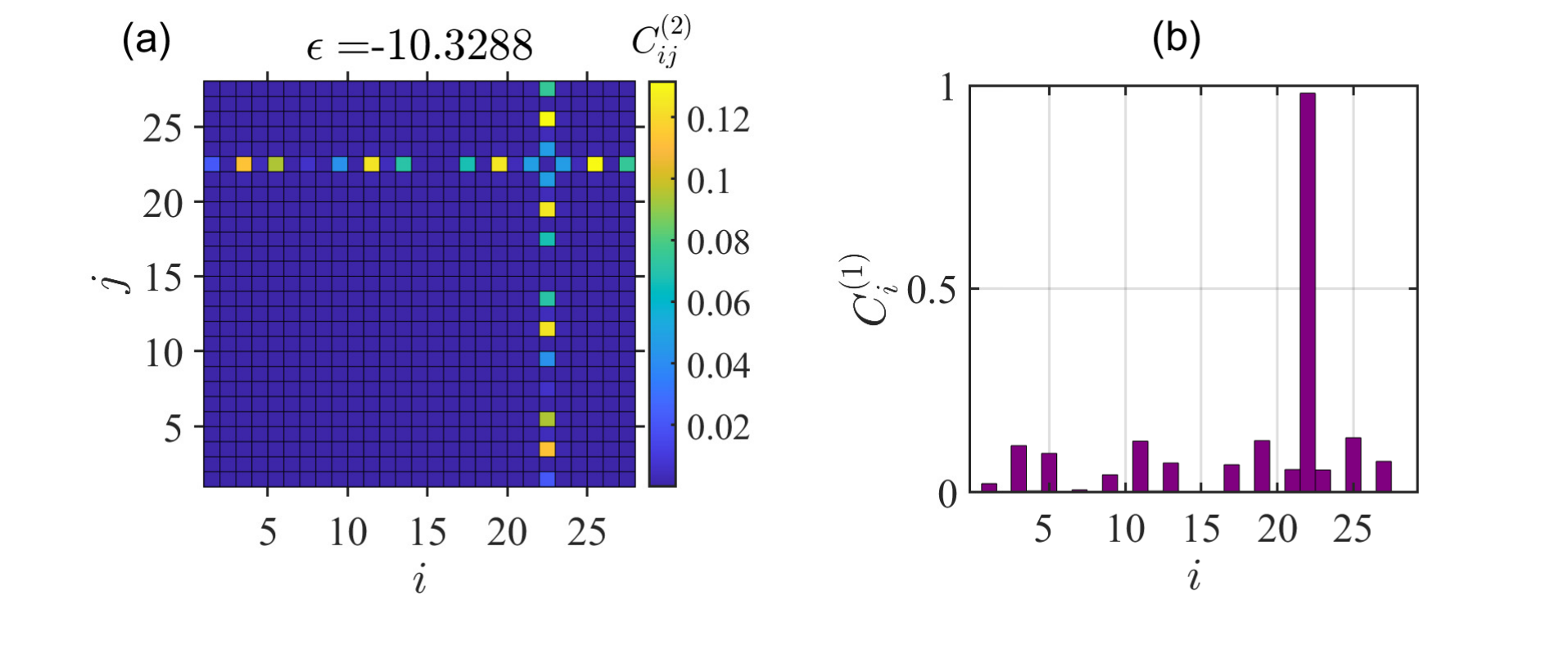}
   \caption{\label{Fig_CORR2}(a) and (b) Second-order and first-order correlation functions of one self-localized state in two-particle system. Our calculation is performed under $U=20J$, $p/q=1/4$, $V=10J$, $\xi=-\pi \beta$, and $L=28$.}
 \end{figure}

In the two-particle system, some of its eigenstates also have the features of self-localization.
In such states, one of the particles is extended in spatial space whereas another particle is strongly localized; see Fig.~\ref{Fig_CORR2}. 
Most of the partially localized states $\Psi^{(2)}$ can be well approximated by the ansatz, $\Psi^{(2)}_{i,j}=\varphi^{(1)}_{i}\chi^{(1)}_{j}+\varphi^{(1)}_{j}\chi^{(1)}_{i}$. 
We can obtain the localized state $\chi^{(1)}$ from singular value decomposition (SVD) of $\Psi^{(2)}$. Like the independent ALL in three-particle system, the localized particle can localize randomly at some lattice sites.
 \subsection{A. Proportion of self-localized states}
 \label{sec:diagram_two}
\begin{figure}[!htp]
  \includegraphics[width=0.45\columnwidth]{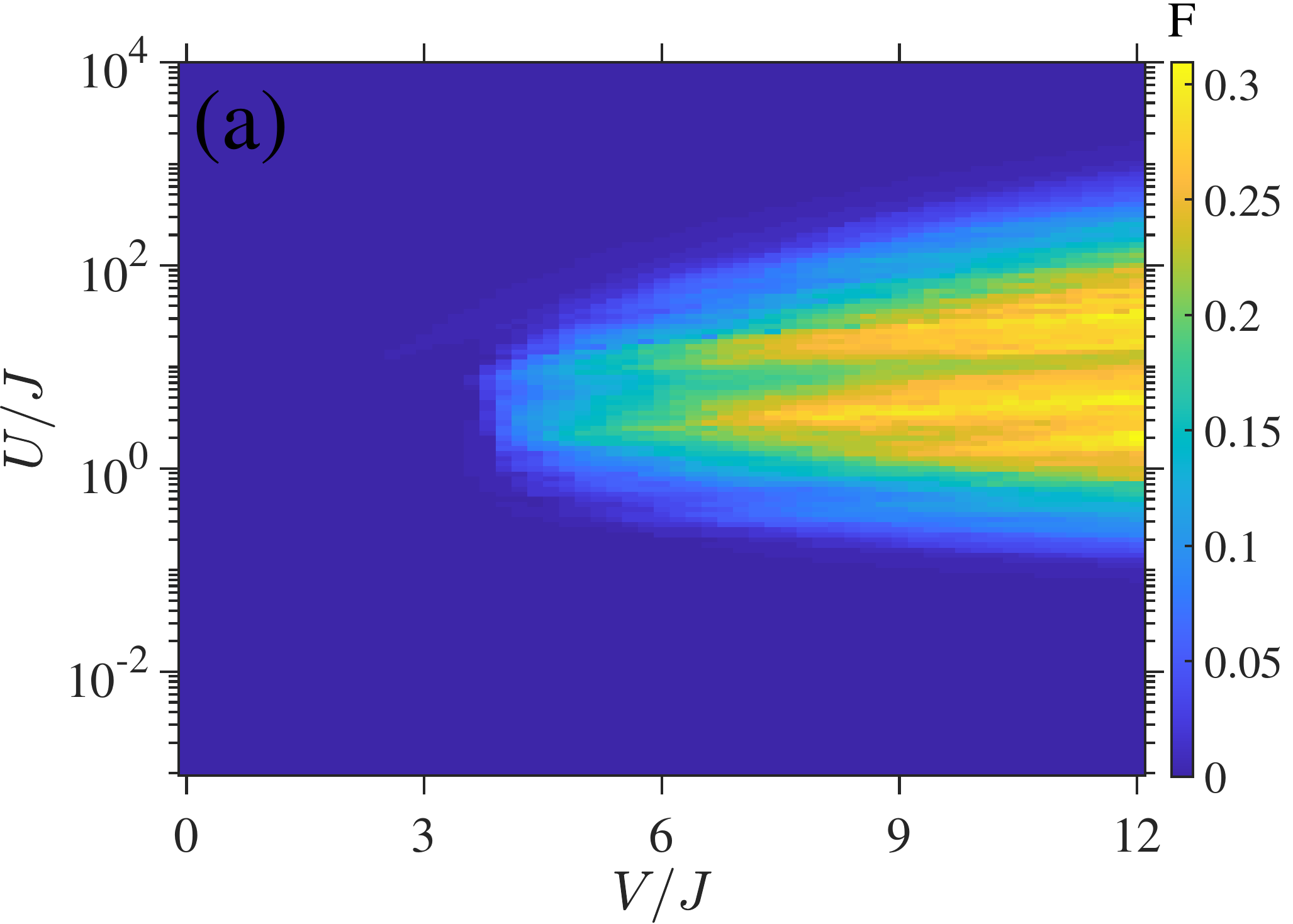}
   \includegraphics[width=0.425\columnwidth]{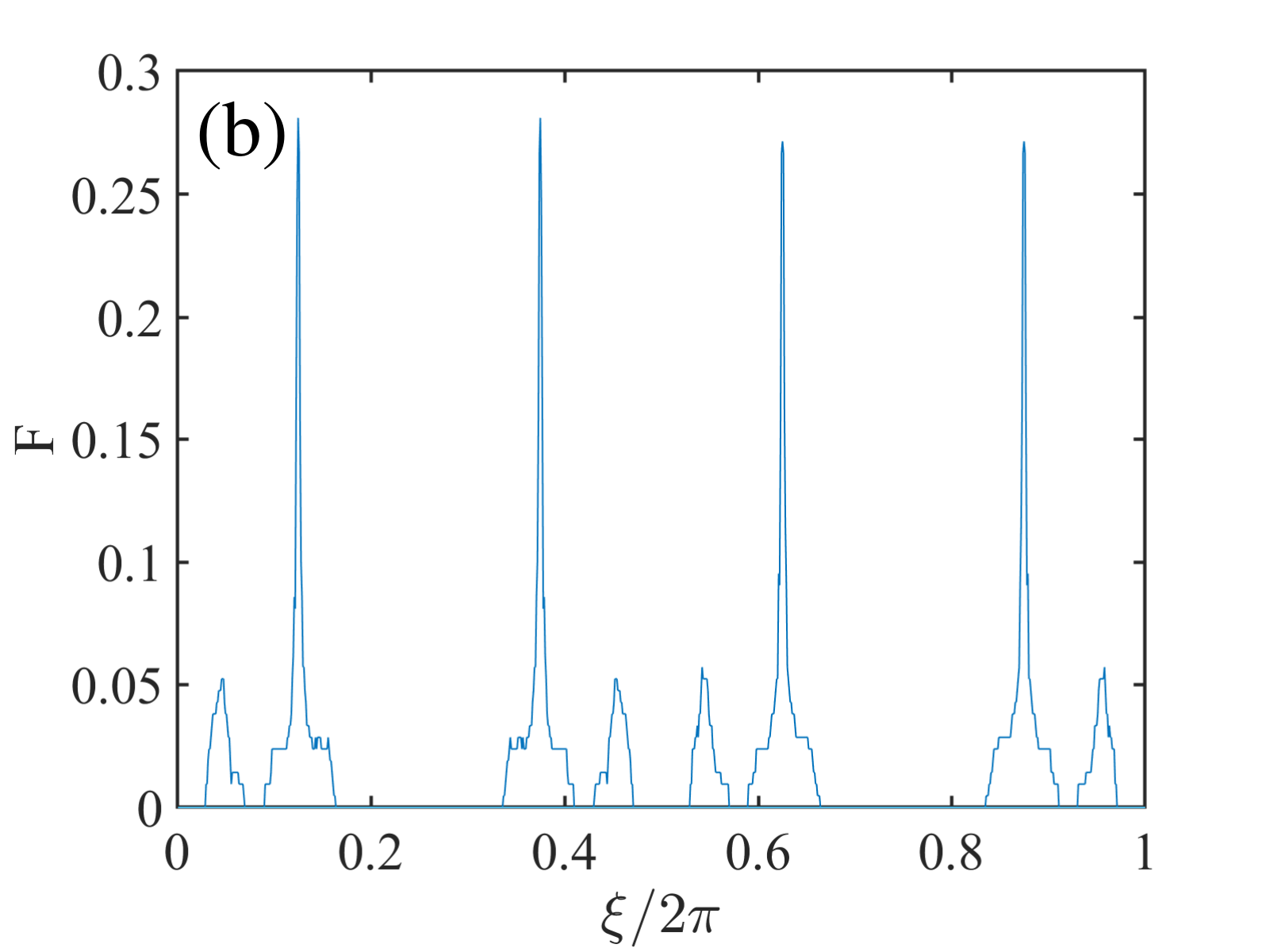}

   \caption{\label{Fig_Diagram2}(a) Fraction of the localized states that can be explained by the effective Hamiltonian as functions of the interaction strength $U/J$ and the depth of the onsite potential $V/J$. 
   Our calculation is performed under $p/q=1/4$, $\xi=-\pi \beta$, and $L=20$.  (b) Fraction of the localized states as a function of modulated phase $\xi$. Our calculation is performed under $p/q=1/4$, $U/J=20$, $V/J=10$, $L=20$.}
 \end{figure}

 Fraction of localized states that can be well explained by the effective Hamiltonian, depends on ($V/J,U/J$) and $\xi$ are shown in Figs.~\ref{Fig_Diagram2} (a) and (b), respectively. 
We find that the self-localized states exist in a broad range of parameters.  
In the two-boby case, the self-localization is a co-effect of interaction and spatial modulation.
Besides, the fraction of self-localized states is also significantly affected by the modulated phase $\xi$, show in Fig.~\ref{Fig_Diagram2} (b).  
The large fraction of self-localized states appears at $\xi_n= \pi/4+n\pi/2$ ($n=0,1,2,3$).
We pick out the localized states by requiring that (i) the sum of the two leading singular values in SVD is larger than 0.8; (ii) the inverse participation radio (IPR) of $\chi$ is larger than 0.8 and that of $\varphi$ is smaller than 0.3; and (iii)  $\langle\tilde{\Psi}^{(2)}|\Psi^{(2)}\rangle>0.9$ where $\tilde{\Psi}^{(2)}$ is the reconstructed two-particle state obtained by product of $\chi$ (one eigenstate of the effective Hamiltonian) and the extended state $\varphi$. 
Since the edge states sharing the same form with self-localized states exist in the non-interacting system, we also exclude their contribution in the calculations. 
 
  \subsection{B. Level spacing statistics}
\begin{figure}[!htp]
    \includegraphics[width=0.59\columnwidth]{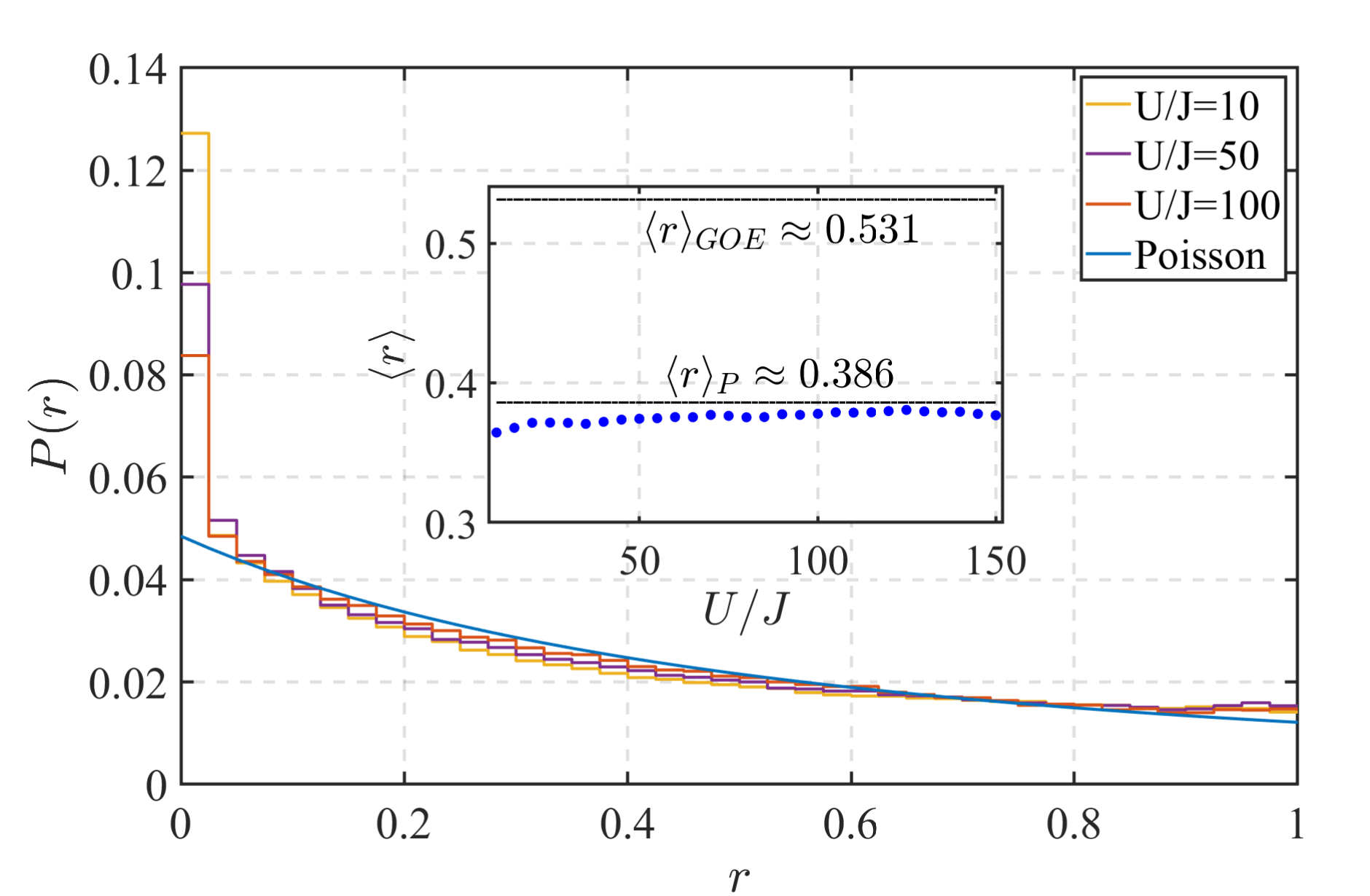}
   \caption{\label{Fig_S3}Energy level spacing distribution $P(r)$ for different interaction strength $U/J$. The inset figure shows the average of the adjacent gap radio $\langle r \rangle$ for different $U/J$. Four states at the band edges are included in calculating $P(r)$ and $\langle r \rangle$. The parameters are the same as those for Fig.~2 in the main text. }
 \end{figure}

 Before calculating the statistics of the energy level, we need to single out the partially self-localized states from all eigenstates. 
 The screening conditions are almost the same as Sec. 2A, except that the IPR of the extended state is smaller than 0.06 in this part.
 We set the parameters as $V/J=10$, $L=64$, $p/q=1/4$, and the ranges of the modulated phase $\xi/2 \pi$ are $\xi_n\pm 0.002\pi$ ($n=0,1,2,3$) with sample interval $5\times 10^{-5}$. 
 The ranges of the modulated phase $\xi$ are near the four peaks displayed in Fig.~\ref{Fig_Diagram2} (b).
 The parameters are the same as those we considered in the main text.
 Under one set of parameters, we can obtain plenty of two-particle self-localized states. 
 For each state, we obtain one extended state and we construct one effective Hamiltonian (one sample).
 We change the modulated phase $\xi$ and we acquire lots of samples. 
 We calculated the statistics of level spacing distribution $P(r)$ and the average level spacing $\langle r\rangle$ for the effective Hamiltonians; see Fig.~\ref{Fig_S3}. 
 Since the radio of the adjacent energy gap for the two states near each energy gap is quite close to 0, there is a significant peak at $r \to 0$. 
Since there are two energy gaps in the single-particle Hamiltonian ($p/q=1/4$), there are 4 fixed values very close to 0 for each sample in our problem. 
If the size of the system is very large, the peaks at $r \to 0$ will decrease to the value predicted by Possion statistics.
However, the size of the system in numerical calculations is limited to $L=64$, the significant peak is visible in our results. 
We also calculate $P(r)$ and $\langle r\rangle$ in which the four states are excluded; see Fig.2 of the main text.
Although there are some differences in the quantities by excluding the four states at $r \to 0$, the main features of Possion statistics maintain the same.

 \subsection{C. Standing waves and hinds for state preparation }
\begin{figure}[!h]
    \includegraphics[width=0.93\columnwidth]{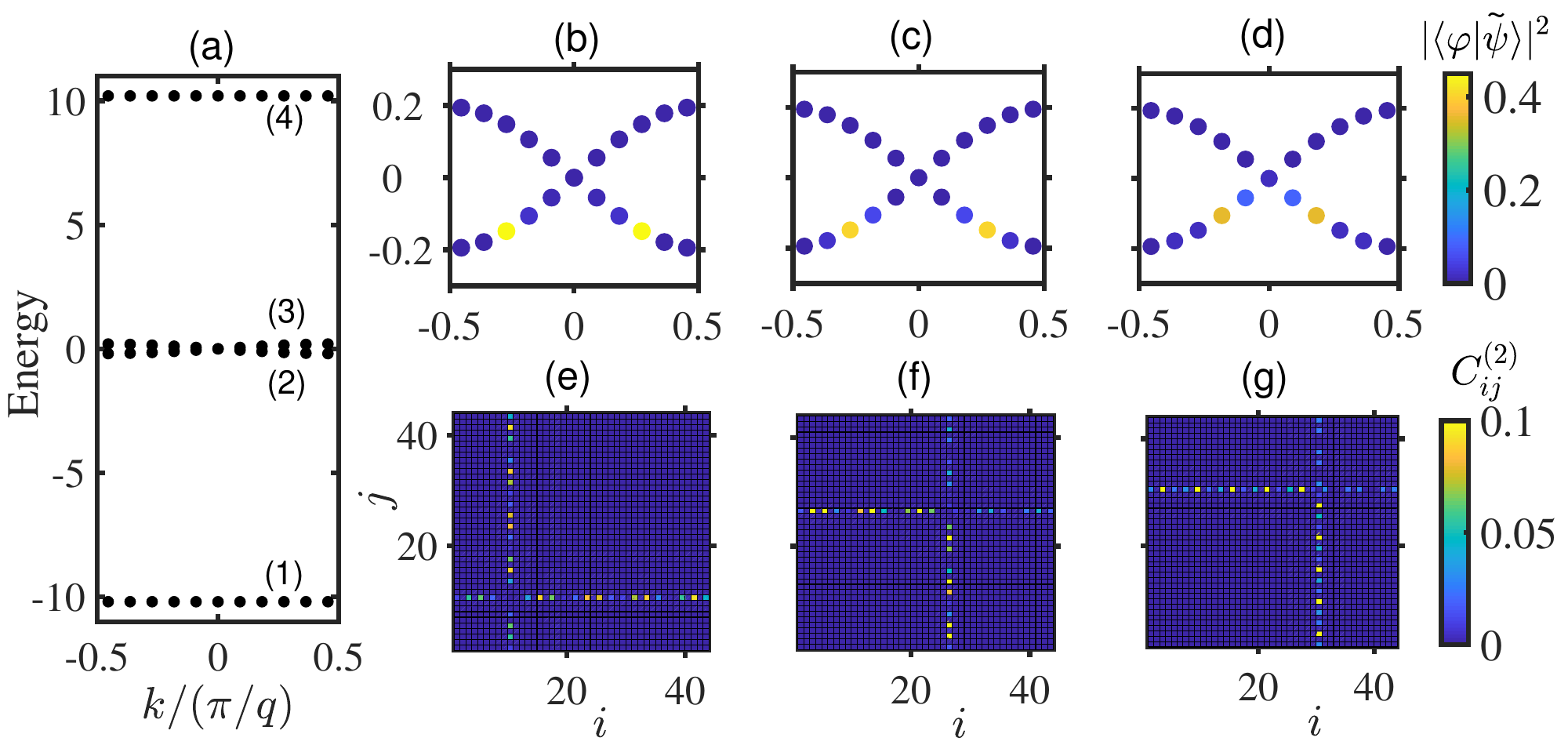}
   \caption{\label{Fig_pro}(a) Energy band for the single-particle system under periodic-boundary condition. (b)-(d) The projections onto the Bloch states for the extended states $\varphi^{(1)}$ of the two-particle states, whose second-order correlation functions are shown in (e)-(g), respectively.
  }
 \end{figure}

We project the extended state $\varphi^{(1)}$ onto the Bloch states. 
 Bloch states are the eigenstates $\tilde{\psi}^{(1)}$ of the single-particle Hamiltonian under periodic boundary condition. Considering $ \ V=10J, U=20J, L=44, p/q=1/4, \xi=-\pi \beta$, as shown in Fig.~\ref{Fig_pro}(a), the single-particle system contains four bands and the energies of two bands in the middle are very close.
Compared with the two middle bands, the top and bottom bands are more flat.  
Figs.~\ref{Fig_pro} (b)-(d) present the projections onto the Bloch states of three $\varphi^{(1)}$ for the two-particle states in Figs.~\ref{Fig_pro} (e)-(g), respectively. 
Most of the projections are distributed in the middle two bands. 
The figures demonstrate that the extended states $\varphi$ are the equal-probability superposition of Bloch states with momentum $k$ and $-k$, that is, they are standing waves.

These features motivate us to propose a dynamical scheme for observing interaction-induced Anderson-like localization.
The key is to use quantum walks to form multiple standing waves.
A particle in a single site is a superposition of Bloch states with many pairs of momenta $k$ and $-k$.
In quantum walks before hitting the boundaries, a particle with opposite momentum moves in different directions, and hence the wavefunctions with opposite momenta do not overlap in space.
To construct standing waves, we need to rely on the reflection of the particle by boundaries so that the momenta change directions.
By repeating reflections, the wavefunctions with opposite momenta are extended over the whole space and they form standing waves.
It will become clearer in Sec. 3F.

\section{S3. Three-particle system}
\subsection{A. Localized position}
\begin{figure}[!htp]
    \includegraphics[width=0.9\columnwidth]{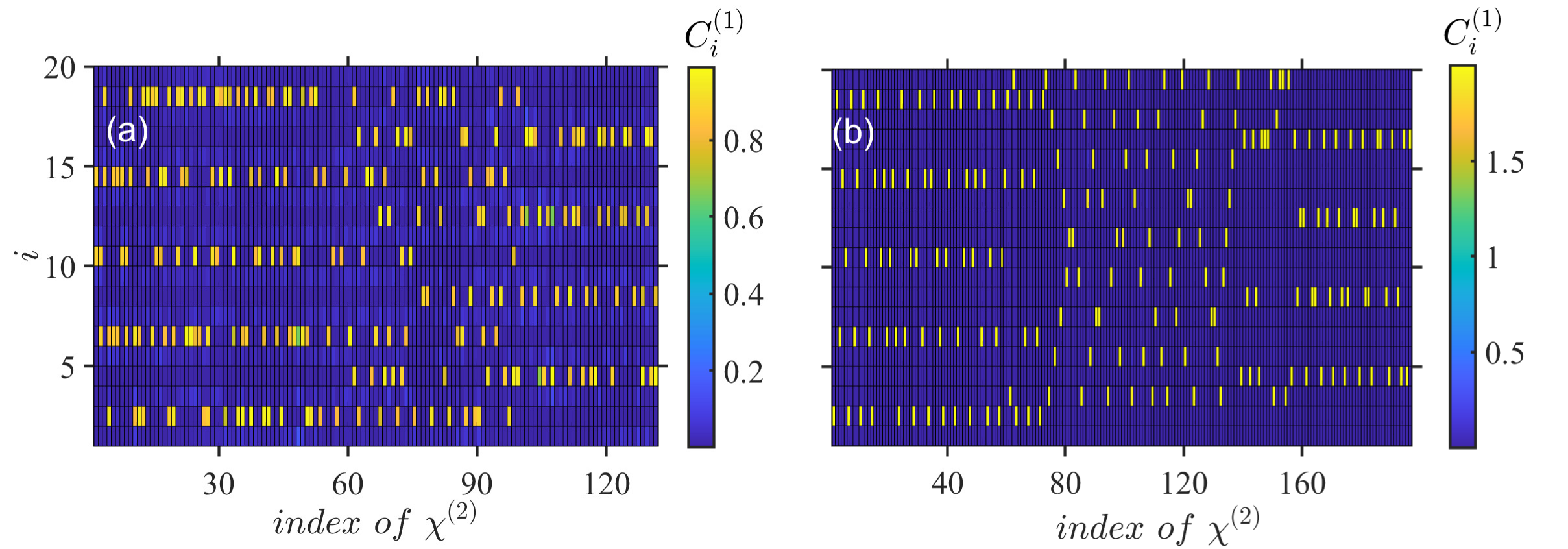}
   \caption{\label{Fig_LP}(a) and (b) First-order correlation functions for $\chi^{(2)}$ for selected three-particle states of the independent ALL and correlated ALL, respectively. We set parameters as $V=10J$, $p/q=1/4$, $L=20$, $\xi=-\pi\beta$, $U=20J$ for (a) and $U=50J$ for (b). }
 \end{figure}
For the three-particle system, the two kinds of localized states can be described by the ansatz 
  \begin{equation}
     \Psi^{(3)}_{i,j,k}=\varphi^{(1)}_{i}\chi^{(2)}_{j,k}+\varphi^{(1)}_{j}\chi^{(2)}_{i,k}+\varphi^{(1)}_{k}\chi^{(2)}_{i,j}.
  \end{equation}
We consider $V=10J$, $U=20J$, $p/q=1/4$, $L=20$, $\xi=-\pi\beta$.
Following the method in Sec. 3B, we acquire the wavefunction $\chi^{(2)}$ of the self-localized states from SVD.
 For independent ALLs, we give density distributions of $\chi^{(2)}$ for different self-localized states in Fig.~\ref{Fig_LP}(a).
  The maximal density of $\chi^{(2)}$ is randomly localized at some of the sites in the lattice. 
  For correlated ALLs, when the interaction strength is weak, the conclusion is similar to the independent ALLs, but it behaves differently from the independent ALLs when $U$ is larger. 
When the interaction strength is increased to $U=50J$, the localization position can be any site of the lattice; see Fig.~\ref{Fig_LP}(b). 
The results are similar with the case of $U=50J$, when $U$ is further increased, e.g., $U=100J,\ 1000J,\ 10000J$.

\subsection{B. Derivation of the ansatz}
\label{sec:svd}
 In this section, we try to illustrate how to obtain the ansatz in the main text $\psi_{i,j,k}\approx\varphi_{i}\chi_{j,k}+\varphi_{j}\chi_{i,k}+\varphi_{k}\chi_{i,j}$ from SVD. For the three-particle eigenstate $\Psi^{(3)}=\sum_{i,j,k}\psi_{i,j,k}|i,j,k\rangle$, we reshape the third-order tensor $\psi_{i,j,k}$ with dimension $ L \times L \times L$ into a matrix $\tilde{\psi}_{i,r}$ with dimension $ L \times L ^2$, where $r=\left(j-1\right)\times L+k$. And then we perform SVD on the matrix, $\tilde{\psi}_{i,r}=\sum_{m,n}
S_{i,n}D_{nm}W_{m,r}$, where the dimensions of the three matrices $S, D$ and $W$ are $L\times L$, $L\times L^2$, and $L^2 \times L^2$, respectively.
We will discuss the independent and correlated ALLs separately due to the difference in SVD between them.


For the independent ALL, the result of SVD for the eigenstate is dominated by three terms, 
\begin{equation}
    \tilde{\psi}_{i,r}\approx D_{11}S_{i,1}W_{1,r}+D_{22}S_{i,2}W_{2,r}+D_{33}S_{i,3}W_{3,r}.
\end{equation}
Since the three singular values are very close $\left(D_{11}\approx D_{22}\approx D_{33}\right)$, we ignore the overall coefficient and then yield 
\begin{equation}
    \tilde{\psi}_{i,r} \approx S_{i,1}W_{1,r}+S_{i,2}W_{2,r}+S_{i,3}W_{3,r}.
\end{equation}
The numerical calculation shows that one of the three single-particle states $S_{i,1},S_{i,2},S_{i,3}$ is always strongly localized in spatial space, and we specify the localized state as $S_{i,1}^{(l)}$.  For the other two states,
we can obtain one localized state and one extended single-particle state after linear combinations of them, $\frac{1}{\sqrt{2}}(S_{i,2}+S_{i,3})$ and $\frac{1}{\sqrt{2}}(S_{i,2}-S_{i,3})$. 
In addition, we perform the parallel transformation, $\frac{1}{\sqrt{2}}(W_{2,r}+W_{3,r})$ and $\frac{1}{\sqrt{2}}(W_{2,r}-W_{3,r})$. 
For the two single-particle states $\frac{1}{\sqrt{2}}(S_{i,2}+S_{i,3})$ and $\frac{1}{\sqrt{2}}(S_{i,2}-S_{i,3})$, we define the state with larger IPR as ${s}^{(l)}$, and the other extended state as $s^{(f)}$, and $w^{(l,f)}(w^{(l,l)})$ corresponding to ${{s}^{(l)}}(s^{(f)})$.
To make the symbols uniform, we rewrite $S_{i,1}^{(l)}$ as $\tilde{s}^{(l)}_i$ and $W_{1,r}^{(l,f)}$ as $\tilde{w}_r^{(l,f)}$.
Therefore, the state $\Psi^{(3)}$ can be expressed as
\begin{equation}
\tilde{\psi} \approx \tilde{s}_{i}^{(l)}\tilde{w}_{r}^{(l,f)}+{s}^{(l)}_{i}{w}^{(l,f)}_{r}+s^{(f)}_{i}w^{(l,l)}_{r}.
\end{equation}
$w^{(l,l)}$ is a localized two-particle state.
For the other two states, $w^{(l,f)}$ and $\tilde{w}^{(l,f)}$, we reshape the vectors with dimension $L^2$ into a matrix with dimensions $L \times L$. 
Performing SVD on the two states, we find that the singular values of both of them are dominated by two terms, and their values are quite close. Thus, the two-particle states can be well  
described by $w^{(l,f)}_{i,j}=\mu_{i}^{(l)}\mu_{j}^{(f)}+\mu_{j}^{(l)}\mu_{i}^{(f)}$, also by $\tilde{w}^{(l,f)}_{i,j}={\nu}_{i}^{(l)}{\nu}_{j}^{(f)}+{\nu}_{j}^{(l)}{\nu}_{i}^{(f)}$. 
The state $\tilde{\psi}_{i,j,k}$ can be expressed as
\begin{equation}\label{17}
 \tilde{\psi}_{i,j,k}\approx \tilde{s}_{i}^{(l)}\left( {\nu}_{j}^{(l)}{\nu}_{k}^{(f)}+{\nu}_{k}^{(l)}{\nu}_{j}^{(f)}\right)+{s}_{i}^{(l)}\left( {\mu}_{j}^{(l)}{\mu}_{k}^{(f)}+{\mu}_{k}^{(l)}{\mu}_{j}^{(f)}\right)+s^{(f)}_{i}w^{(l,l)}_{j,k}.
\end{equation}
So far, we obtain three extended states, ${s}^{(f)}$, $\mu^{(f)}$, and ${\nu}^{(f)}$. 
For most independent ALL states, the inner products between two of the three are close to $1$ after normalization, that is, ${s}^{(f)}\approx \mu^{(f)} \approx {\nu}^{(f)}$. 
We define the extended single-particle state as $\varphi^{(f)}$, $ \varphi^{(f)}= {s}^{(f)}\approx \mu^{(f)} \approx {\nu}^{(f)} $.
We extract the common extended state $\varphi^{(f)}$ in Eq.~\eqref{17} and define the summation of the leaving part as $\chi^{(l)}$,
\begin{equation}\label{eqq}
 \chi_{j,k}^{(l)}= \left(\tilde{s}_{j}^{(l)} {\nu}_{k}^{(l)}+\tilde{s}_{k}^{(l)} {\nu}_{j}^{(l)}\right)+\left({s}_{j}^{(l)} {\mu}_{k}^{(l)}+{s}_{k}^{(l)} {\mu}_{j}^{(l)}\right)+w_{j,k}^{(l,l)},
\end{equation}
 which is a two-particle localized state. Then the three-particle state can be expressed as 
 \begin{equation}
     \tilde{\psi}_{i,j,k}=\varphi^{(f)}_{i}\chi_{j,k}^{(l)}+\varphi^{(f)}_{j}\chi_{i,k}^{(l)}+\varphi^{(f)}_{k}\chi_{i,j}^{(l)}.
 \end{equation}

For the correlated ALL, the result of SVD for the eigenstate is dominated by two terms, 
\begin{equation}
\tilde{\psi}\approx D_{11}S_{i,1}W_{1,r}+D_{22}S_{i,2}W_{2,r}.
\end{equation}
The numerical calculation shows that one of the two single-particle states $S_{i,1},S_{i,2}$ is always strongly localized in spatial space, and we specify the localized state as $S_{i,1}^{(l)}$ and the extended state as $S_{i,2}^{(f)}$.
$W_{2,r}^{(l,l)}$ is a localized two-particle state.
For $W_{1,r}^{(l,f)}$, we reshape the vector with dimension $L^2$ into a matrix with dimensions $L \times L$. 
For simplicity, we rewrite $W_{1,r}^{(l,f)}$, $W_{2,r}^{(l,l)}, S_{i,1}^{(l)},S_{i,2}^{(f)}$  as $w^{(l,f)}$, ${w}^{(l,l)},s^{(l)},s^{(f)}$. 
Performing SVD on $W_{1,r}^{(l,f)}$, we find that the singular value of the state is dominated by two terms and their values are quite close. Thus, the two-particle states can be well  
described by $w^{(l,f)}_{i,j}=\mu_{i}^{(l)}\mu_{j}^{(f)}+\mu_{j}^{(l)}\mu_{i}^{(f)}$. Then the state $\tilde{\psi}_{i,j,k}$ can be expressed as
\begin{equation}\label{eqq}
 \tilde{\psi}_{i,j,k}\approx D_{11}s_{i}^{(l)} \left( {\mu}_{j}^{(l)}{\mu}_{k}^{(f)}+{\mu}_{k}^{(l)}{\mu}_{j}^{(f)}\right)+D_{22}s_{i}^{(f)} {w}^{(l,l)}_{j,k}.
\end{equation}
So far we have obtained two extended states $s^{(f)}$ and $\mu^{(f)}$. 
For most of the correlated ALL states, the inner products between the two extended states are close to $1$ after normalization, that is, $s^{(f)}\approx \mu^{(f)}$. 
We define the extended single-particle state as $\varphi^{(f)}=  s^{(f)}\approx \mu^{(f)}$ and the summation of the left parts as $\chi^{(l)}$,
\begin{equation}\label{eqq}
 \chi_{j,k}^{(l)}= D_{11}\left({s}_{j}^{(l)}{\mu}_{k}^{(l)}+{s}_{k}^{(l)}{\mu}_{j}^{(l)}\right)+D_{22}w_{j,k}^{(l,l)},
\end{equation}
 which is a two-particle localized state. Then the three-particle state can be expressed as
 \begin{equation}
{\psi}_{i,j,k}=\varphi^{(f)}_{i}\chi_{j,k}^{(l)}+\varphi^{(f)}_{j}\chi_{i,k}^{(l)}+\varphi^{(f)}_{k}\chi_{i,j}^{(l)}.
 \end{equation}

\subsection{C. Validity of the general method}
\begin{figure}[!h]
    \includegraphics[width=0.98\columnwidth]{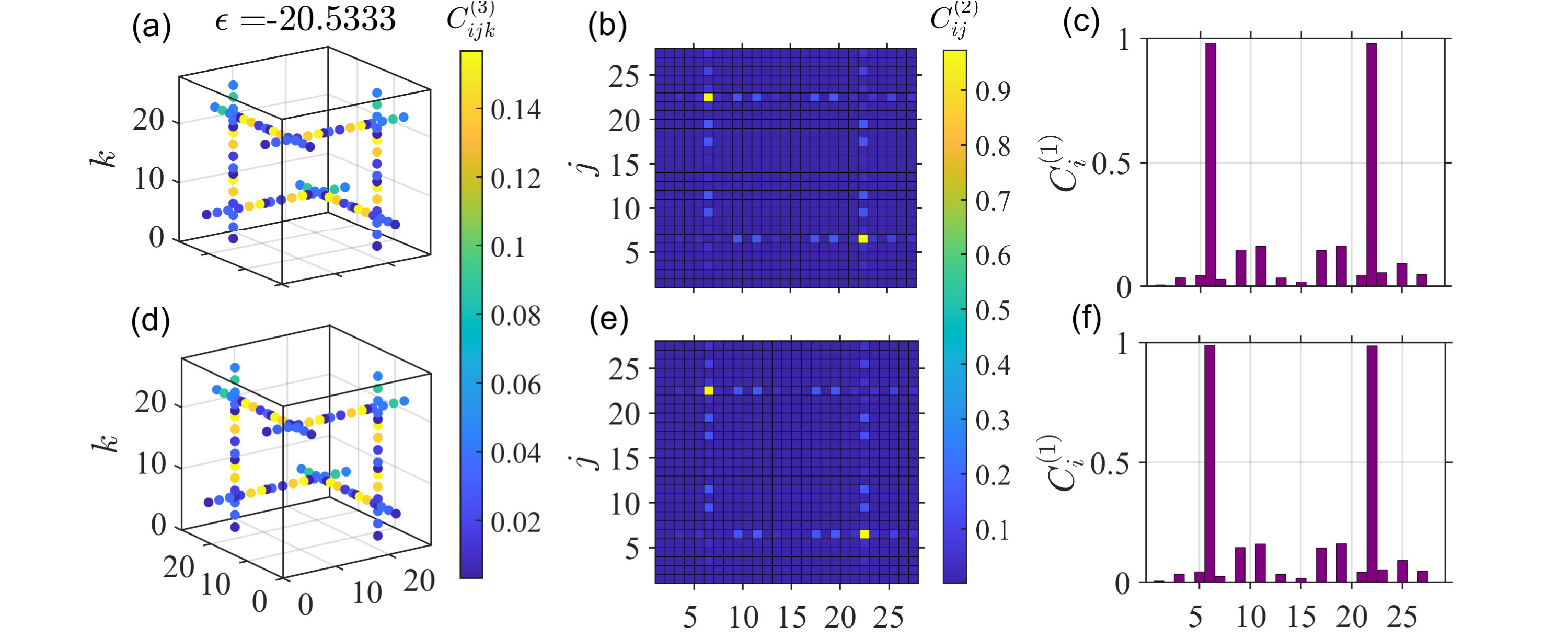}
       \includegraphics[width=0.98\columnwidth]{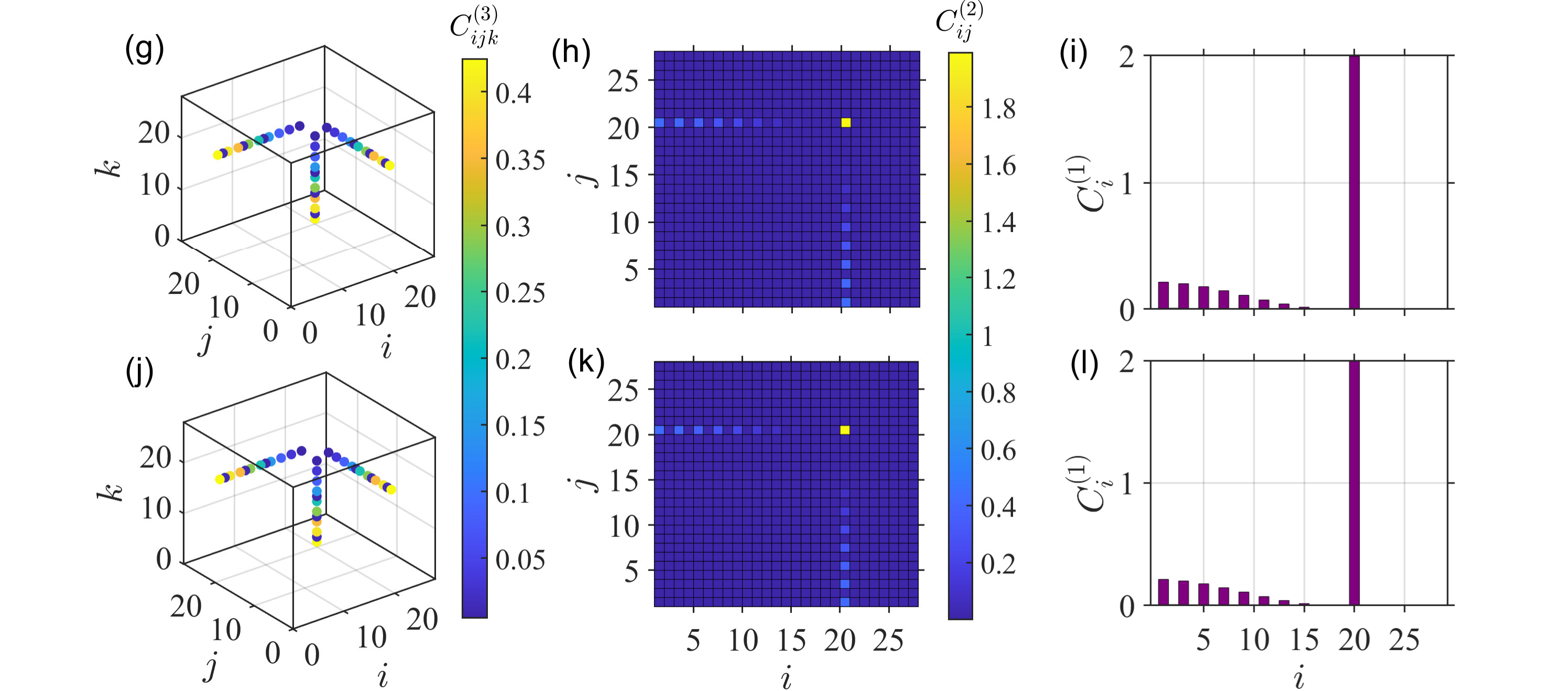}
   \caption{\label{Fig_VS}(a)-(c) Correlation functions of the third-, second- and first-order for independent ALL, obtained by exact diagonalization.  (d)-(f) Similar to (a)-(c), but acquired from the effective model. (g)-(i) Correlation functions of the third-, second- and first-order correlation functions for correlated ALL, obtained by exact diagonalization. (j)-(l) Similar to (g)-(i), but acquired from the effective model. Calculations are performed under $ U=20J, p/q=1/4, V=10J, L=28, \xi=-\pi\beta$.
  }
 \end{figure}

\begin{figure}[!h]
    \includegraphics[width=0.8\columnwidth]{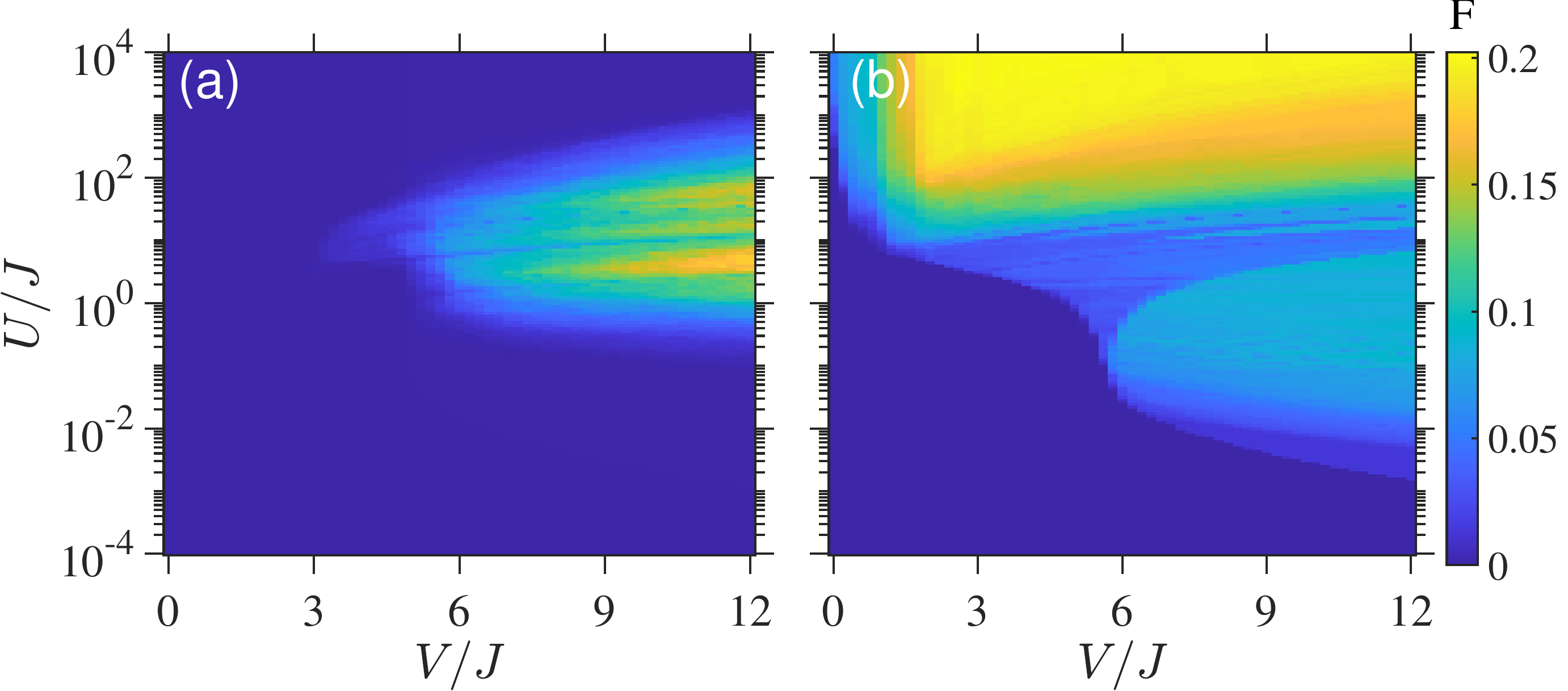}
   \caption{\label{Fig_diagram3_sup}(a) and (b) Fraction of independent ALLs and correlated ALLs,  which can be well described by the effective Hamiltonian, as
a function of ($U/J$, $V/J$), respectively. 
Calculations are performed under $ p/q=1/4,L=16, \xi=-\pi\beta$.
  }
 \end{figure}

In this section, we intend to examine the validity of the effective model. An intuitive approach is to compare the distribution of the correlation functions between the state obtained by the effective model and the ones obtained by the exact diagonalization. 
We take the case of three-particle localized states as an example. For the independent ALL, the correlation functions of an eigenstate $\Psi^{(3)}$ acquired by exact diagonalization are plotted in Figs.~\ref{Fig_VS}(a)-(c). 
The correlation functions of the reconstructed states $\tilde{\Psi}^{(3)}$ acquired by the semi-analytical method are displayed in Figs.~\ref{Fig_VS}(d)-(f). 
It shows that the results calculated by the two methods are almost the same. 
This is consistent with the result that the fidelity between them is close to 1.  
For the correlated ALL, the correlation functions of one eigenstate obtained by exact diagonalization and effective model are shown in Figs.~\ref{Fig_VS}(g)-(i) and (j)-(l), respectively. 
Since the number of localized states is large, it is not convenient to compare the distribution of the correlation functions of each state. 
It is feasible to calculate the validity between $\Psi^{(3)}$ and $\tilde{\Psi}^{(3)}$ of each localized state under different sets of parameters to examine the validity of the effective model. 
The proportion of localized states that can be well explained by the effective Hamiltonian depends on the parameters $U/J$ and $V/J$; see Fig.~\ref{Fig_diagram3_sup}. 
Apart from the three conditions for screening self-localized states mentioned in the text, we add one more condition $|\langle \Psi|\tilde{\Psi}\rangle|>$0.9 to select states that can be well described by the general method.
Compared with Figs.~3(d) and 3(h) in the main text, we can find that most of the self-localized states can be well captured by the effective Hamiltonian.

\subsection{D. Simple lattice case: $q=1$}
\begin{figure}[!htp]
    \includegraphics[width=0.93\columnwidth]{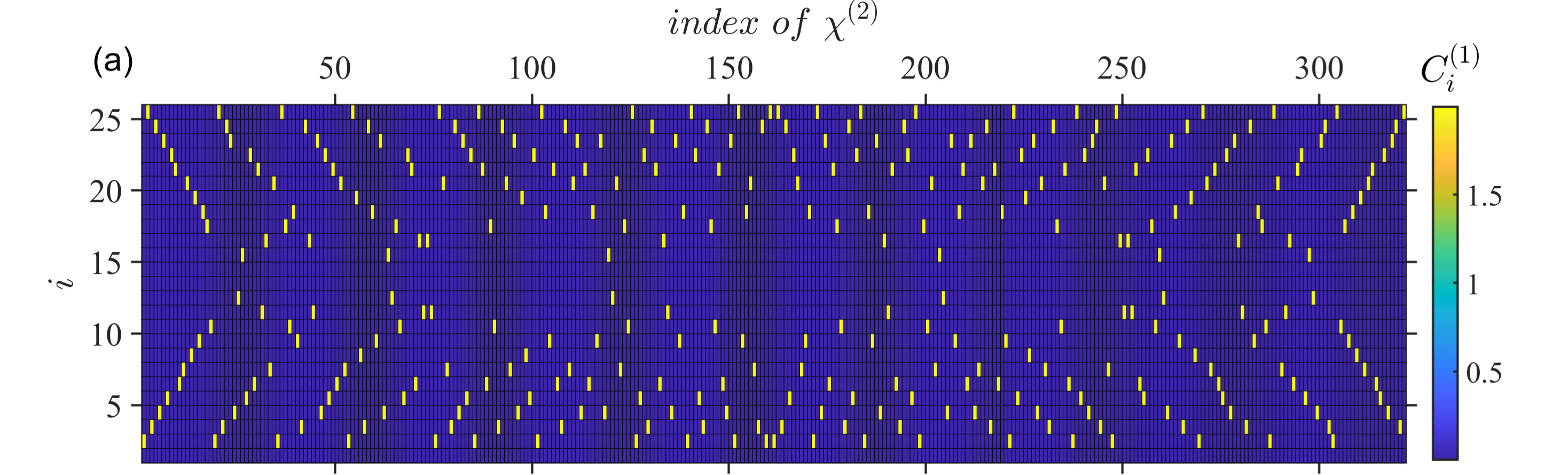}
      \includegraphics[width=0.93\columnwidth]{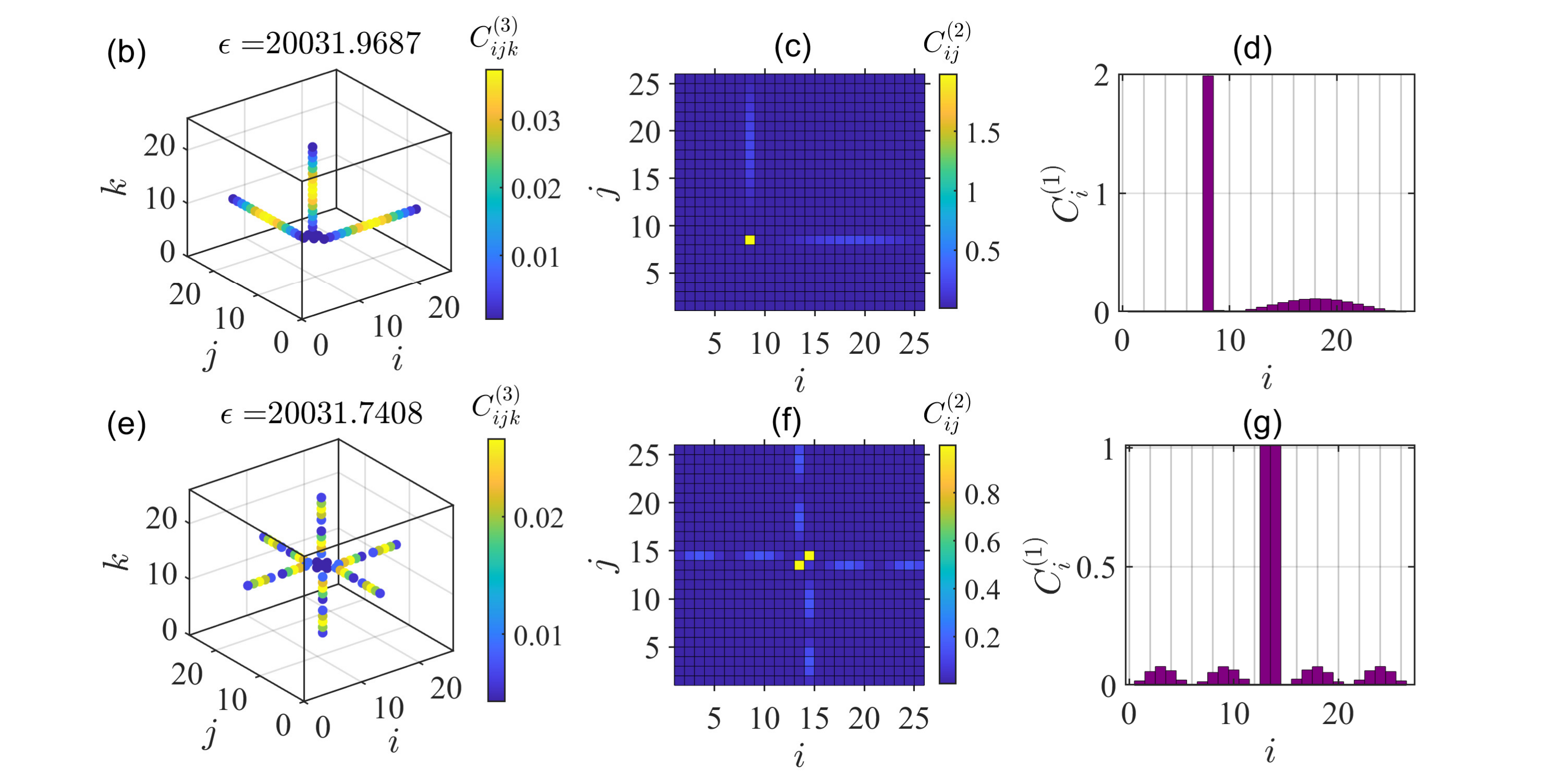}
   \caption{\label{Fig_BH}(a) First-order correlation functions of $\chi^{(2)}$ for selected three-particle self-localized states $\Psi^{(3)}$. (b-d) Three-, second- and first-order correlation functions for one self-localized three-particle state, in which the bound dimer locates on only one site. (e-g) similar to (b-d), but for one self-localized three-particle state, in which the bound dimer locates on two nearest sites. Calculations are performed under $U=20000J, V=0, L=28, \xi=-\pi\beta$.
  }
 \end{figure}

Interestingly, the localization phenomenon of correlated ALL still exists in the absence of the modulated potential.
We choose $ U=20000J$, $V=0$, $L=28$, $\xi=-\pi \beta$, and select the correlated ALL like  Fig.~\ref{Fig_BH}(b) according to the screening conditions mentioned above. 
Figs.~\ref{Fig_BH}(b)-(d) show the third-, second-, and first-order correlation functions for one of these states.
The selected states can be well described by the ansatz $\Psi^{(3)}_{i,j,k}=\varphi^{(1)}_{i}\chi^{(2)}_{j,k}+\varphi^{(1)}_{j}\chi^{(2)}_{i,k}+\varphi^{(1)}_{k}\chi^{(2)}_{i,j}$. 
Following the method described in Sec. 3B, we obtain one $\chi^{(2)}$ for each selected self-localized $\Psi^{(3)}$.
The first-order correlation function $C_{i}^{(1)}$ of $\chi^{{(2)}}$ for each selected $\Psi^{(3)}$ is shown in Fig.~\ref{Fig_BH}(a). 
It gives the localized position of the dimer for the selected three-particle states. 
It seems that the bound dimer can localize on each of the sites except for the two sites in the middle. 
In fact, there exist localized states in the two middle sites; see Figs.~\ref{Fig_BH}(e)-(g). 
Unlike the states in Figs.~\ref{Fig_BH}(b)-(d), the two bound particles populate two nearest sites instead of one site.
Moreover, the SVD feature of the states shown in Fig.~\ref{Fig_BH}(e) is different from that of the state in Fig.~\ref{Fig_BH}(b). The singular values are no longer dominated by two terms and the state can not be captured by the ansatz above, so that the states like Fig.~\ref{Fig_BH}(e) are not included in Fig.~\ref{Fig_BH}(a). 

\subsection{E. States with one localized particle}
\begin{figure}[!htp]
\includegraphics[width=1\columnwidth]{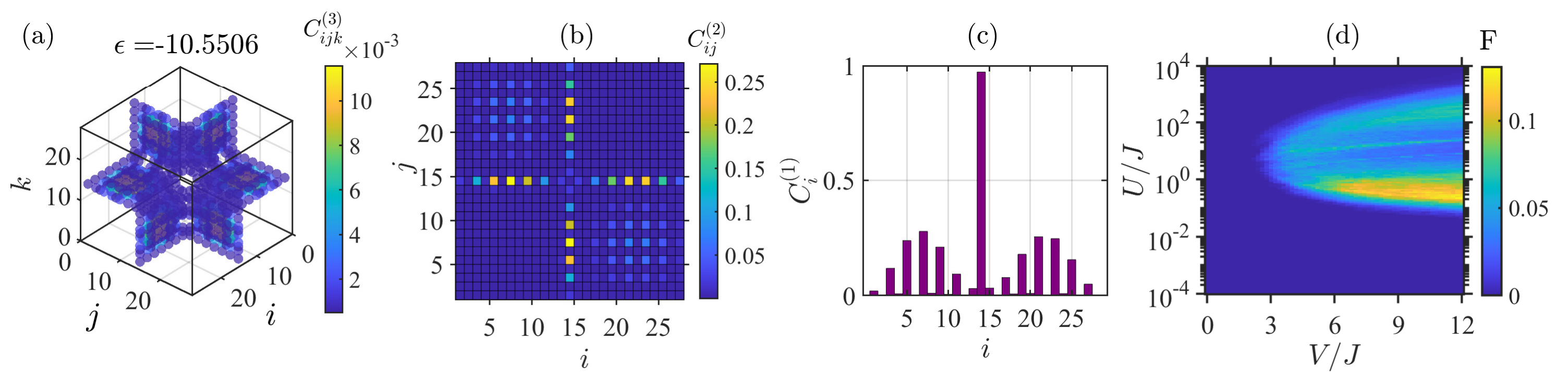}
\caption{\label{Fig_Oneloc}(a-c) Third-order, second-order and first-order correlation functions of one three-particle state which contains one localized particle. (d) Fraction of such kind of self-localized states in
all eigenstates as a function of ($U/J,V/J$). Calculations are performed under $ U=20J$, $V=10J$, $p/q=1/4$, $\xi=-\pi\beta$, $L=28$ for (a-c), $L=20$ for (d).}
 \end{figure}

For completeness, we also show a different type of self-localized states, that is, one of the particles is strongly localized in spatial space, while the other two particles are extended, as shown in Fig.~\ref{Fig_Oneloc}.
Figs.~\ref{Fig_Oneloc}(a)-(c) give the third-, second-, and first-order correlation functions of one eigenstate, respectively. 
Since there are two extended particles, the third-order correlation function looks like three intersecting planes. 
The feature of SVD for this kind of localized states is also different from the other two configurations (independent and correlated ALLs). 
The singular values are no longer dominated by two terms or three terms. 
 The wavefunction $\Psi^{(3)}$ of this type cannot be captured by the ansatz $\Psi^{(3)}_{i,j,k}=\varphi^{(1)}_{i}\chi^{(2)}_{j,k}+\varphi^{(1)}_{j}\chi^{(2)}_{i,k}+\varphi^{(1)}_{k}\chi^{(2)}_{i,j}$.
 %
The fraction of such eigenstates is relatively large, and its dependence on $U/J$ and $V/J$ is shown in Fig.~\ref{Fig_Oneloc}(d).
%
%
%
Here, we try to single out this kind of states based on the distribution of the first-order correlation function $C_{i}$ of the eigenstates. 
We mark the site with the highest value of $C_{i}$ as $i_{a}$. 
We require that  $|\Psi^3(i_a,j,k)|^2>0.8$ or  $|\Psi^3(j,i_a,k)|^2>0.8$ or $|\Psi^3(j,k,i_a)|^2>0.8$ ($j\neq i_a,k\neq i_a$) as well as $C_{i}<0.4$ ($i\neq i_a$) simultaneously. 
 \subsection{F. Dynamical simulation} \label{Dynamics}
\begin{figure}[!h]
   \includegraphics[width=0.85\columnwidth]{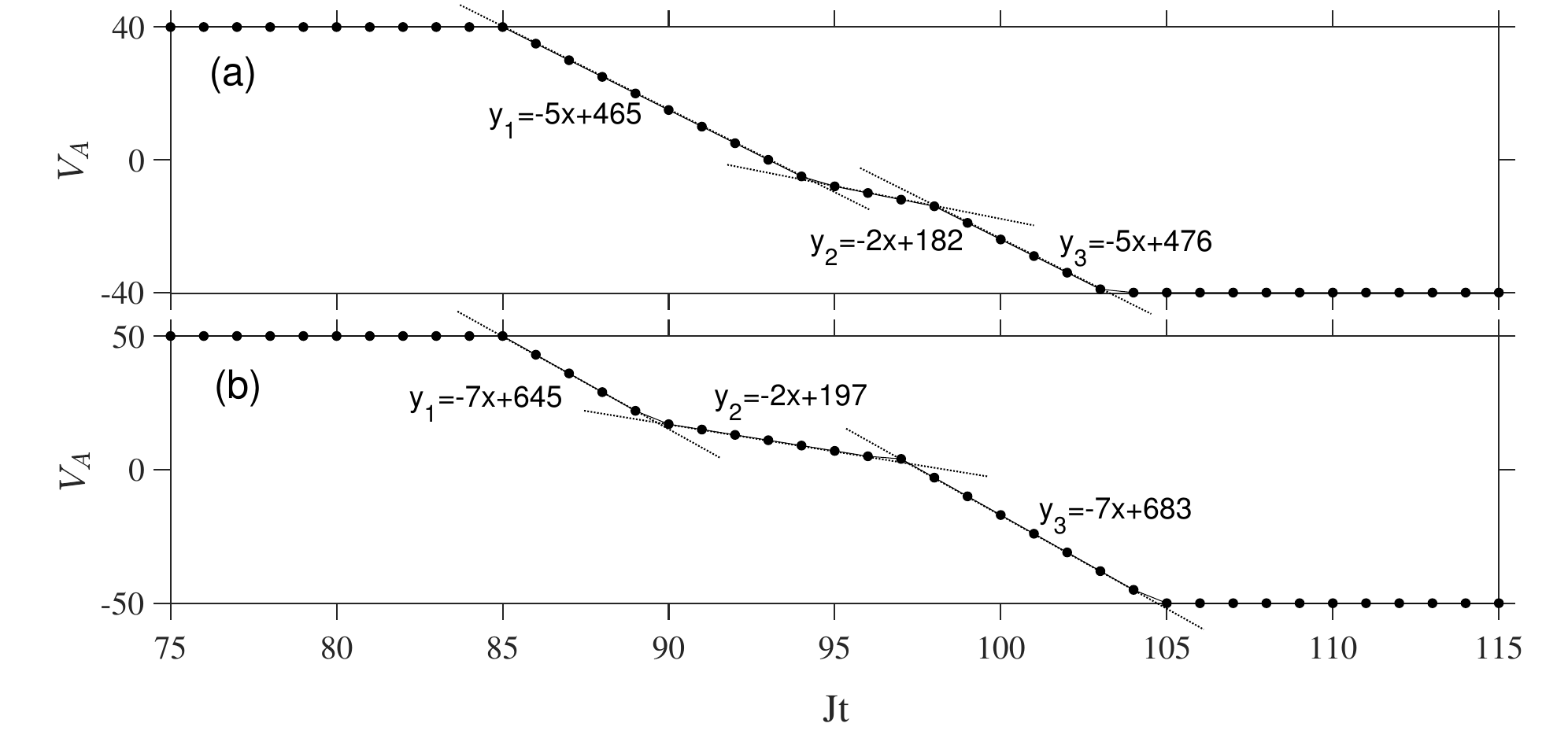}
   \caption{\label{Fig_Potential}(a) and (b) Potential of the auxiliary sites versus time from $Jt=75$ to $Jt=115$ for independent ALL and correlated ALL, respectively. 
  }
 \end{figure}

\begin{figure}[!h]
   \includegraphics[width=0.45\columnwidth]{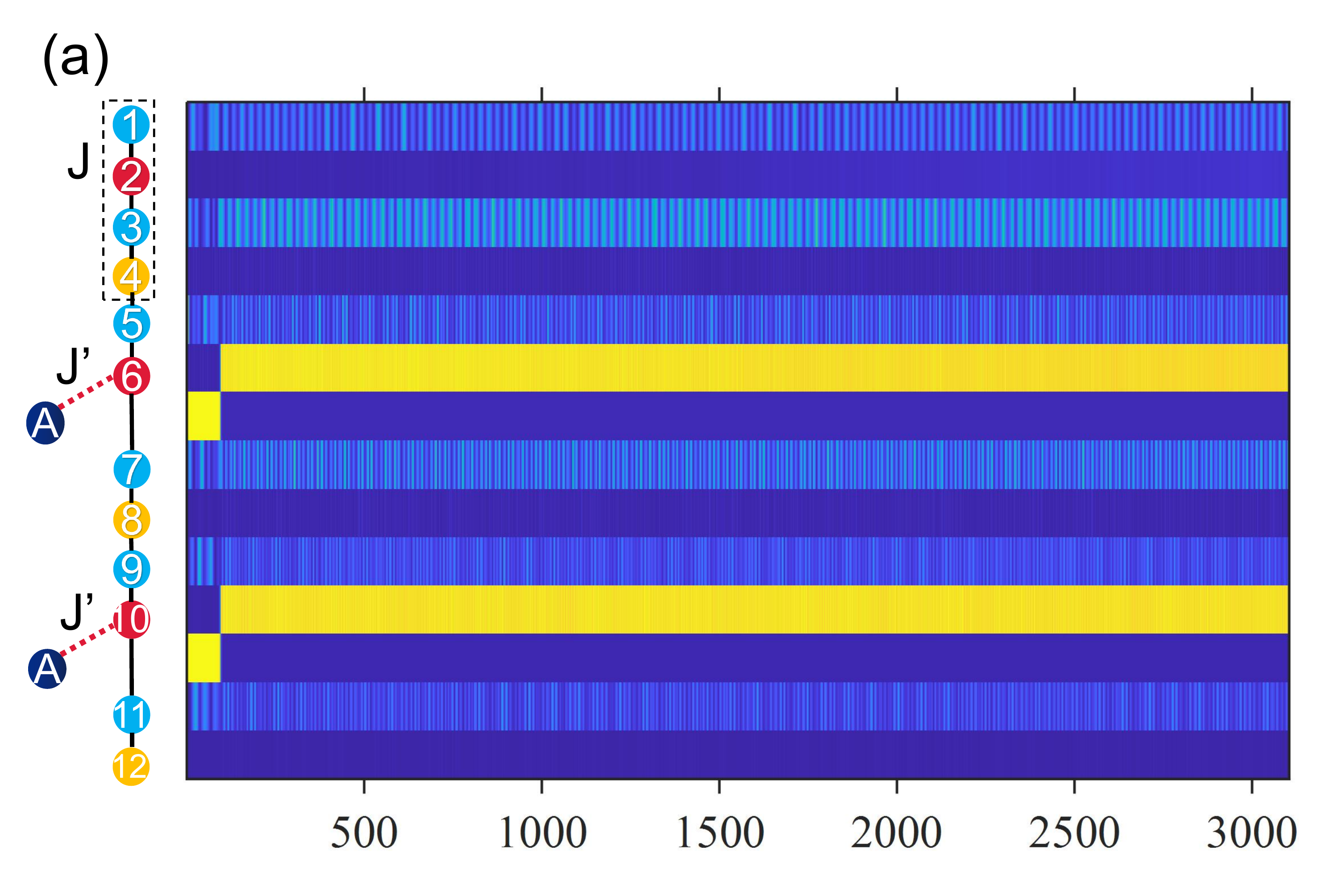}
   \includegraphics[width=0.45\columnwidth]{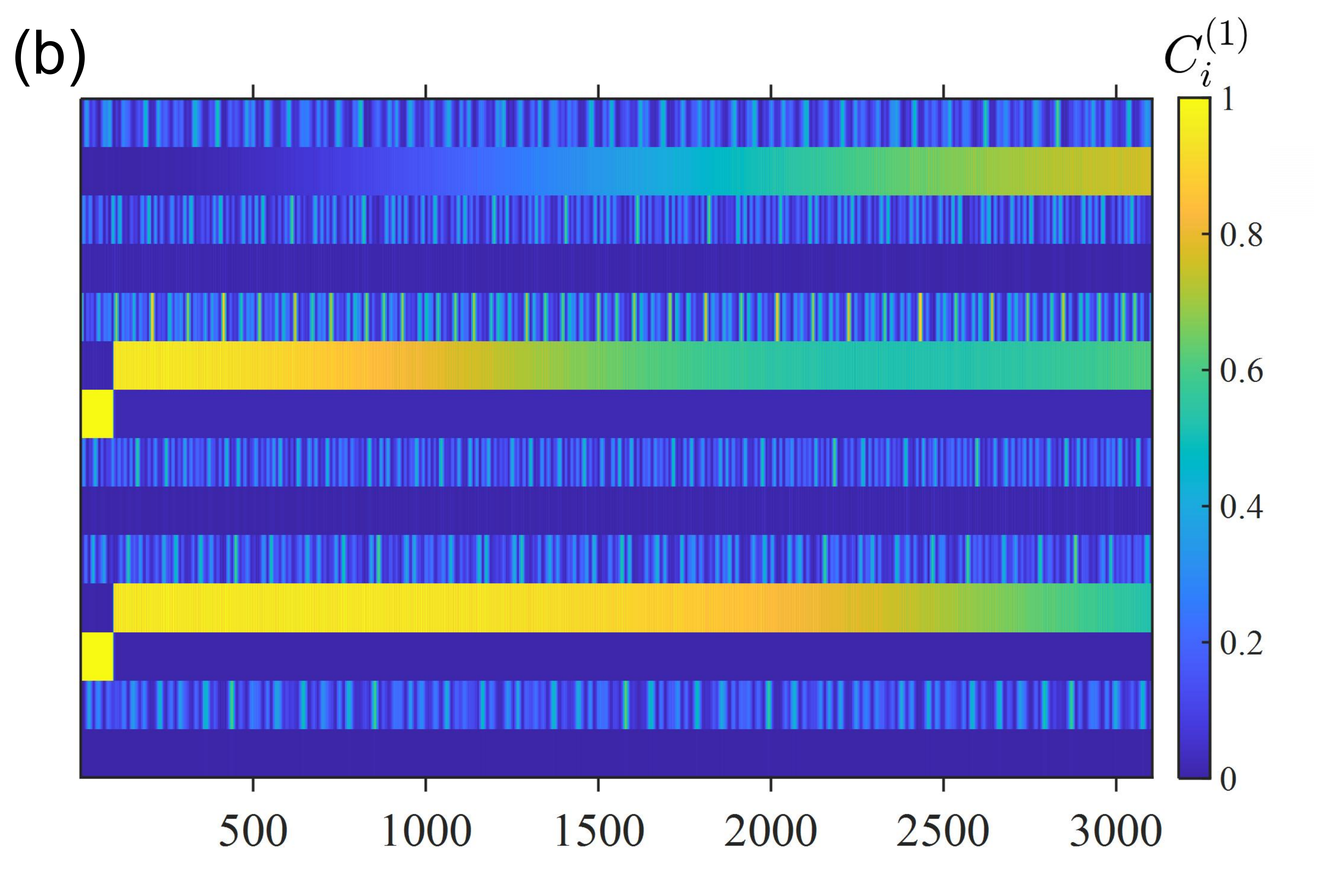}
   \includegraphics[width=0.45\columnwidth]{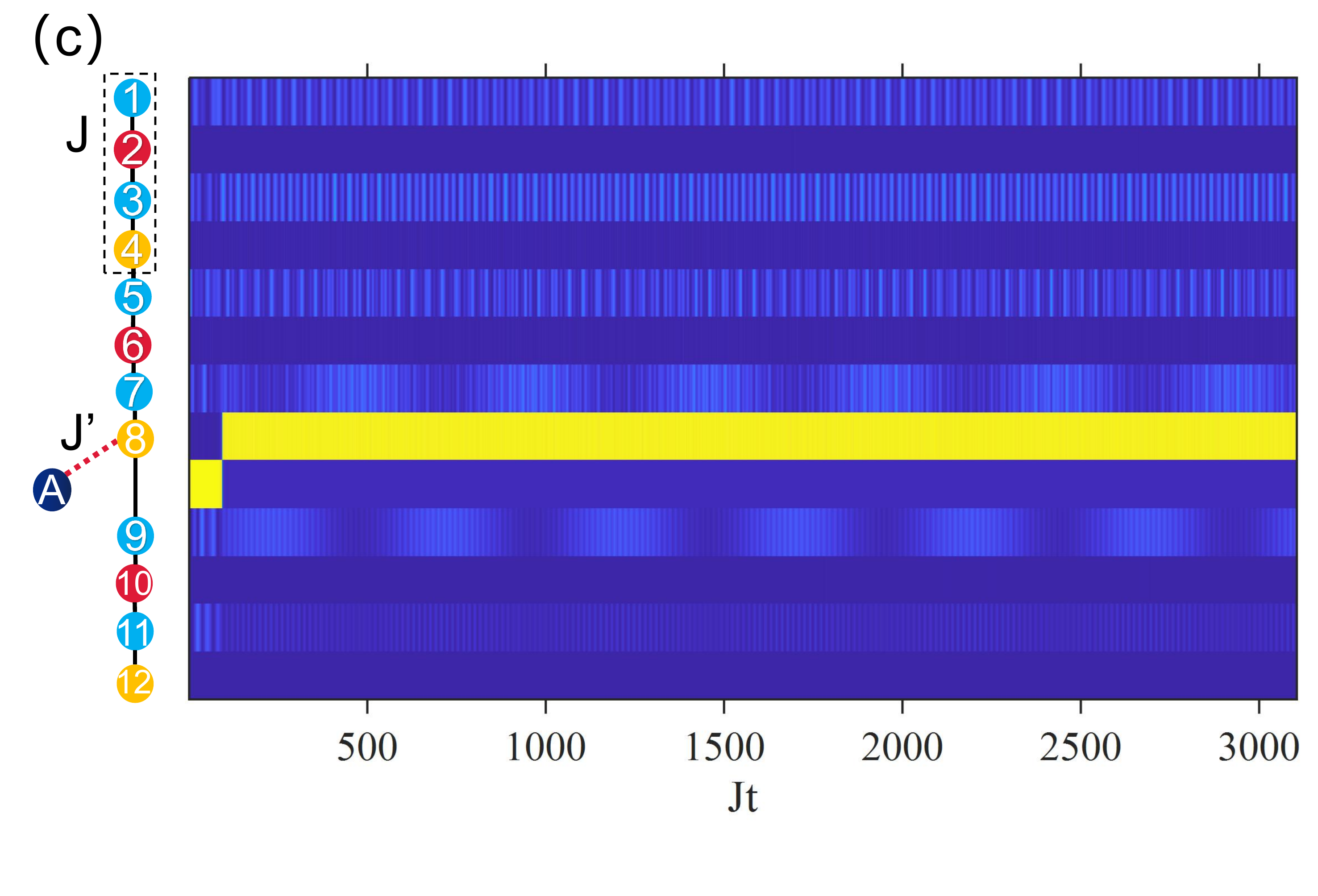}
 \includegraphics[width=0.45\columnwidth]{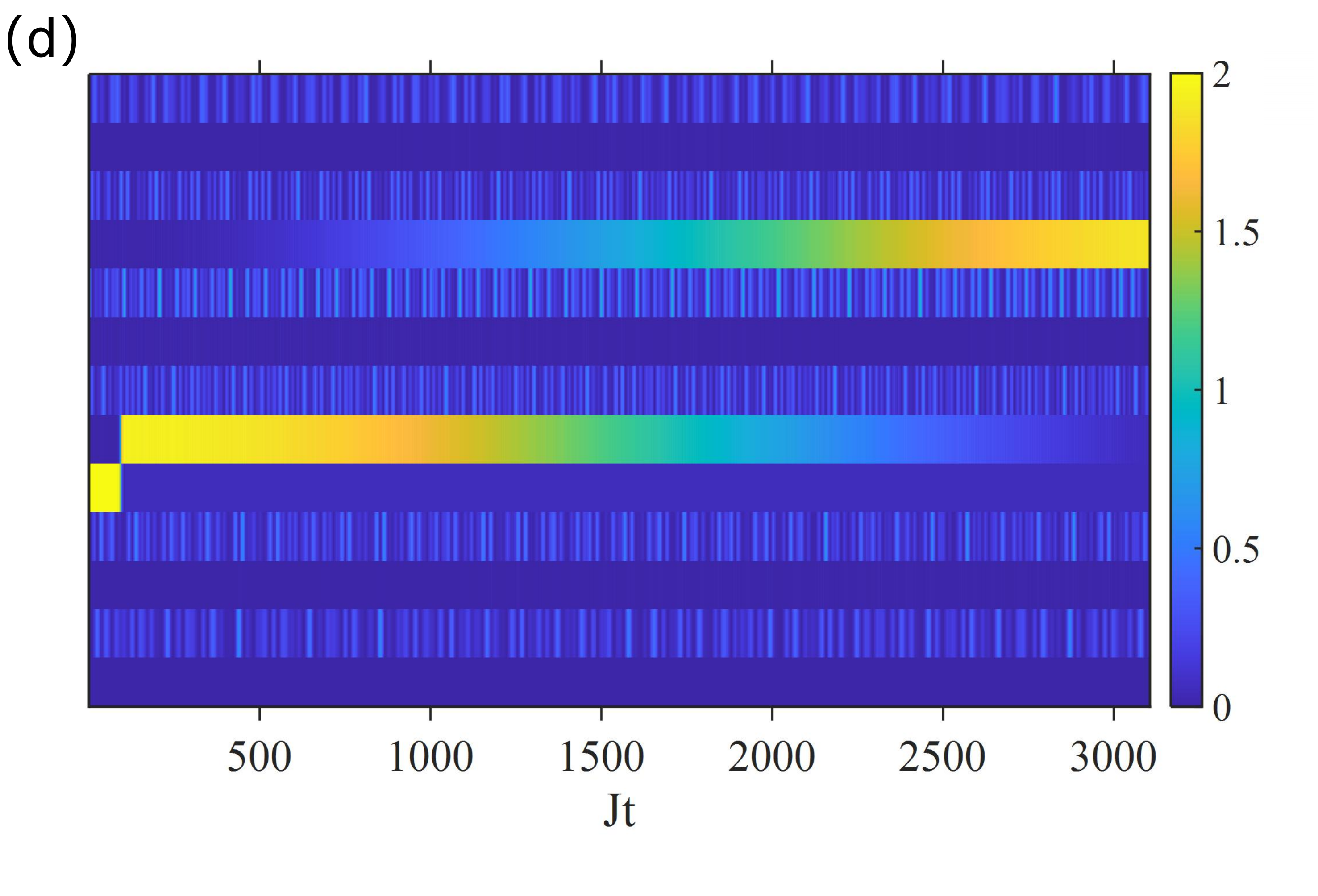}
   \caption{\label{Fig_longtime}(a)-(b) The distribution of first-order correlation functions versus time for the independent ALL with $U=20J$ and $U=0$, respectively. (c)-(d) The same as (a) and (b), but for correlated ALL. The hopping amplitude between the auxiliary sites $A$ and superlattice are chosen as $J'=0$ in processes (i) and (iii).  And we set $J'=1.5J$ in process (ii) for the independent ALL and $J'=4J$ for the correlated ALL. The other parameters are set as $T_{1}=84J/\hbar$, $T_{2}=104J/\hbar$, $T_{3}=3104J/\hbar$, $L=12$, {$V=10J$, $\xi=-\beta\pi$}, and $p/q=1/4$. }
 \end{figure}

We set the parameters as $V=10J$, $L=12$, $\beta=1/4$, $\xi=-\pi \beta$.
For the independent ALL, In the process (i), we prepare one boson at the $5$th site of the superlattice with 3 unit cells containing $12$ sites ($p/q=1/4$). 
The other two bosons are prepared at two different auxiliary sites. The potential depths  of the two auxiliary sites are always the same during the whole process.
While the boson in the superlattice undergoes quantum walks in the lattice for a long time $T_1=84J/\hbar$, the other two bosons are trapped at the auxiliary sites $A$ isolated from the superlattice.
In process (ii), we gradually reduce the potential bias between the auxiliary sites and the superlattice, and turn on the hopping $J'=1.5J$, so that the two trapped particles will be adiabatically and completely transferred to the superlattice at the time $T_2=104J/\hbar$. The strength of the potential of the auxiliary sites $V_A$ versus time is shown in Fig.~\ref{Fig_Potential} (a).
In process (iii), we turn off the hopping $J'=0$ and let the three-particle state undergo a free evolution to time $T_3=3104J/\hbar$; see Figs.~\ref{Fig_longtime} (a) and (b) with $U/J$=20 and $U=0$. 
For the correlated ALL, the strength of the potential of the auxiliary site $V_A$ versus time is shown in Fig.~\ref{Fig_Potential} (b).
The distributions for the first-order correlation function after long-time evolution ($T_{3}=3104J/\hbar$) are shown in Figs.~\ref{Fig_longtime} (c) and (d) with $U/J=20$ and $U=0$, respectively. 
We choose the other parameters the same as the ones in the main text. We find that the localization persists for both the correlated ALL and the independent ALL with $U=20J$. 
As predicted, the localization diffuses for $U=0$ with the other parameters unchanged for both of the two configurations. 
The diffusion of the localization for $U=0$ takes a long time due to the deep onsite potential $V=10J$.

\begin{figure}[!h]
    \includegraphics[width=0.83\columnwidth]{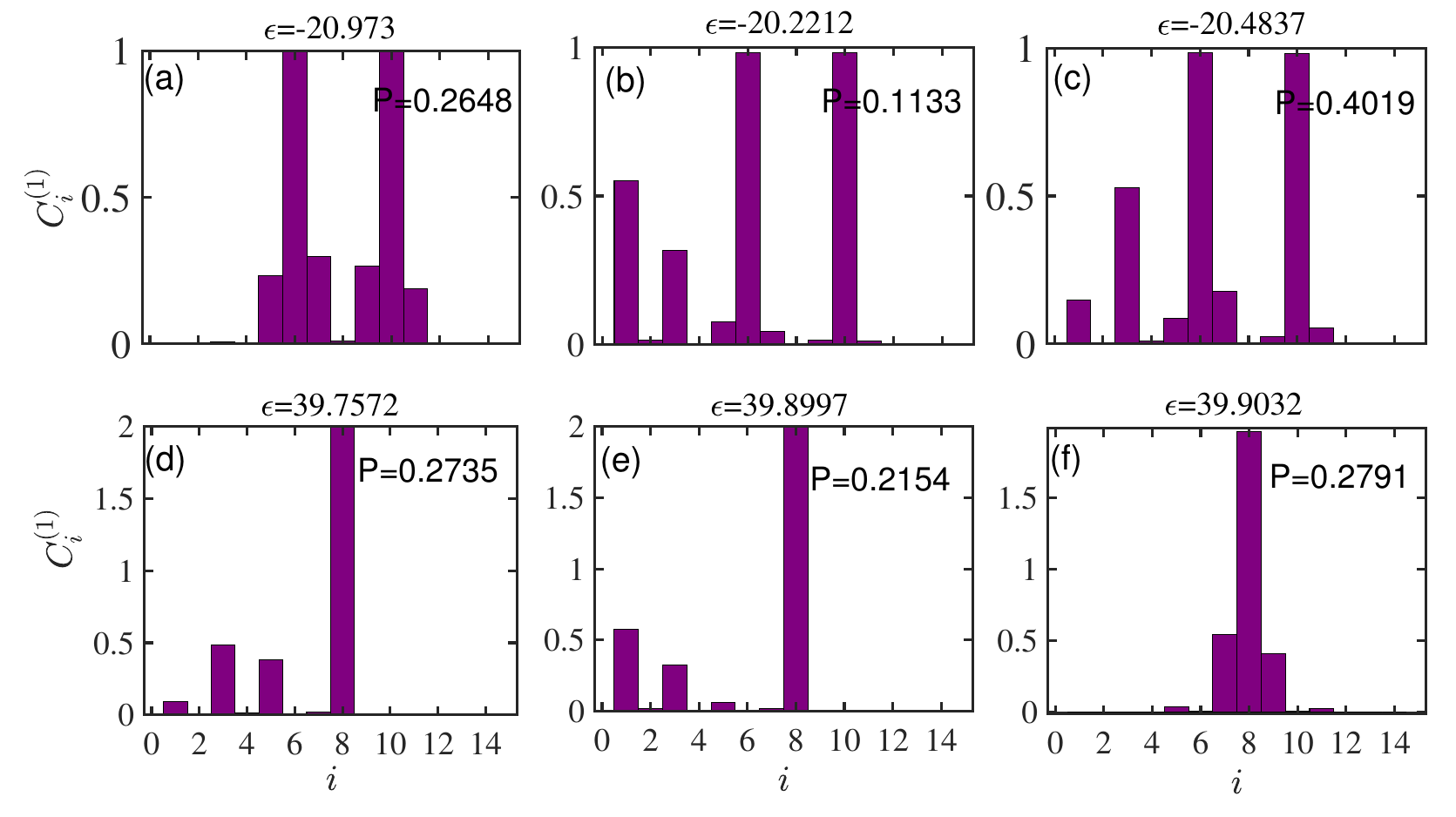}
   \caption{\label{Fig_projection}(a)-(c) First-order correlation functions of $\psi_{t=0}$ with the three largest projection probabilities. $P=|\langle \psi_{t=0}|\Psi\left( T_2 \right)\rangle|^2$.  $\psi_{t=0}$ is the eigenstate of the Hamiltonian at $t=0$. $\Psi_{T_{2}}$ is the prepared three-particle state of the independent ALL at $t=T_2$. (d)-(f) The same as (a-c), but for   correlated ALL.
  }. 
 \end{figure}

For both the correlated ALL and the independent ALL,
we project $\Psi\left( T_2 \right)$ onto eigenstates ($\psi_{t=0}$) of the Hamiltonian at $t=0$. 
For the independent ALL, the first-order correlation functions of $\psi_{t=0}$ with the three largest projection probabilities $P=|\langle \psi_{t=0}|\Psi\left( T_2 \right)\rangle|^2$ are shown in Figs.~\ref{Fig_projection} (a)-(c).
The fourth largest projection probability is $0.0357$ for the localized eigenstate with eigenvalue $\epsilon=-20.3075$. 
The sum of the projection probabilities corresponding to these four states is $0.8157$. 
For the correlated ALL, the first-order correlation functions of $\psi_{t=0}$ with the three largest projection probabilities $|\langle \psi_{t=0}|\Psi\left( T_2 \right) \rangle |^2$ are shown in Figs.~\ref{Fig_projection} (d)-(f). 
In addition to the three states, several localized eigenstates ($\epsilon=40.1080, 40.1622, 40.503$) with relatively smaller projection probability ($P=0.0495,\ 0.067,\ 0.0548$) are not shown. 
The sum of the probabilities corresponding to these six states is $0.9393$. 
Due to the large projection probability for both cases, we can observe independent and correlated self-localization in Figs.~\ref{Fig_longtime}(a) and (c), respectively. 

\end{document}